\documentclass[trackchanges]{aastex7}

\begin{document}

\title{On the origin of super-global spin-up trends of X-ray pulsars in HMXB systems}

\author[0000-0003-1202-9751]{Vitaliy Kim}
\affiliation{Fesenkov Astrophysical Institute, Observatory Street 23, Medeu district, 050020, Almaty, Kazakhstan}
\affiliation{al-Farabi Kazakh National University, al-Farabi Avenue 71, Bostandyk district, 050040, Almaty, Kazakhstan}
\affiliation{Pulkovo observatory, Pulkovskoye shosse 65/1, 196140, Saint-Petersburg, Russia}
\email[show]{kim@fai.kz, ursamajoris1987@gmail.com}

\author[0000-0001-5717-6523]{Yerlan Aimuratov}
\affiliation{Fesenkov Astrophysical Institute, Observatory Street 23, Medeu district, 050020, Almaty, Kazakhstan}
\affiliation{al-Farabi Kazakh National University, al-Farabi Avenue 71, Bostandyk district, 050040, Almaty, Kazakhstan}
\affiliation{International Center for Relativistic Astrophysics Network, Piazza della Repubblica 10, 65122, Pescara, Italy}
\email{aimuratov@fai.kz}

\begin{abstract}
\noindent
Most X-ray pulsars in high-mass X-ray binary (HMXB) systems exhibit both global and local trends of spin acceleration (spin-up) and deceleration (spin-down). Moreover, decades-long monitored objects display even more general spin-up trends in their spin evolution and simultaneously demonstrate a decrease in the orbital period, so-called orbital decay. Although obvious, those general behavior leaves the open question on the energy source and the mechanism of angular momentum transfer that ensures the neutron star's spin acceleration. We hypothesize that the observed spin acceleration of X-ray pulsars in HMXB (RLO and sgXRB) systems results from orbital angular momentum transfer via tidal interactions with their massive companions. To study these phenomena, we sample five well-studied objects with the longest observational history of their spin periods from the population of known Galactic persistent X-ray pulsars in HMXB systems. We obtain the spin frequency change from the observational data and estimate energy loses for orbital decay and pulsars's spin acceleration. We propose a new term of super-global spin-up trends, and show that their observable values do not exceed theoretical upper limits predicted within the scenario of orbital decay due to the tidal interaction between the neutron star and its massive companion.
\end{abstract}

\keywords{High mass X-ray binary stars (733)  --- Neutron stars (1108) --- Pulsars (1306) --- Accretion (14) --- High energy astrophysics (739)}


\section{Introduction} \label{sec:introduction}
X-ray pulsars in high-mass X-ray binary systems (HMXB) are close pairs of stars, where the first component is a neutron star (NS) and the second is a massive star of early O--B or Be spectral class \citep{Neumann...2023AA...677A.134N, Fortin...2023AA...671A.149F, Kim...2023ApJS..268...21K}. These objects can be divided into three groups depending on mass-transfer mode (accretion regime): Roche lobe overflow systems (RLO HMXB pulsars), Be X-ray binary systems (BeXRB pulsars) and pulsars accreting from stellar wind from their giant or super-giant massive companions (sgXRB pulsars) \citep{Chaty...2013AdSpR..52.2132C}. X-ray pulsations of these objects are usually associated with their spin rotation and with a sufficiently strong magnetic field, which affects the nature of the motion of matter near the NS and leads to an inhomogeneous temperature distribution of its surface \citep{Lyne...2012puas.book.....L}. 

The Galactic HMXB population contains at least $82$~known X-ray pulsars \citep{Kim...2023ApJS..268...21K}. These systems exhibit a broad range of spin periods (from $2.76$ to $36$,$200$~s) but a relatively narrow range of measured surface magnetic fields $\sim$($0.86$--$7.80$)$\times10^{12}\,\text{G}$. Their massive companions find themselves in a relatively narrow range of spectral types, O6.5--B3, and a wide range of luminosity classes that extend from main-sequence stars to supergiants \citep{Kim...2023ApJS..268...21K}. The population of X-ray pulsars in HMXB  is divided into two groups: persistent sources and transients. The objects from the first group are characterized by a relatively stable X-ray flux with smooth variations compared to transients demonstrating significant outburst activity \citep{Reig...2011Ap&SS.332....1R, Reig...1999MNRAS.306..100R}. 

Since their discovery, most pulsars have shown variations in their spin periods \citep{Lipunov...1992ans..book.....L, Malacaria...2020ApJ...896...90M}. These spin period variations possess a complex structure on the time scale and can be divided into several smaller components: local and global trends. The fastest changes occur on short-time scales (from several days to several months) and are chaotic when the pulsar spin period undergoes short-term oscillations near the equilibrium period. Such rapid changes are known as local trends of spin-up (or spin-down). These local variations occur against the backdrop of more gradual global trends of spin-up (or spin-down), which last from months to several years \citep{Kim...2023ApJS..268...21K}. 

However, if we analyze the overall dynamics of period changes for objects observed over several decades, it can be noticed that the spin periods of all objects (except for Vela~X-1) are decreasing. To describe this phenomenon, we introduce a new term, the super-global trend of spin-up. We consider the dissipation of the orbital angular momentum (orbital decay) of the HMXB system as a possible energy source for the super-global spin-up of the NS through its interaction with the accretion flow. Conversely, the global spin-up/spin-down trends might be attributed to the drift of the NS's equilibrium period, potentially caused by long-term changes in the stellar wind velocity of the massive companion. On the other hand, local trends appear to stem from minor fluctuations in the accretion rate, likely due to heterogeneity in the stellar wind.

\begin{figure}[ht]
\center{\includegraphics[width=\textwidth]{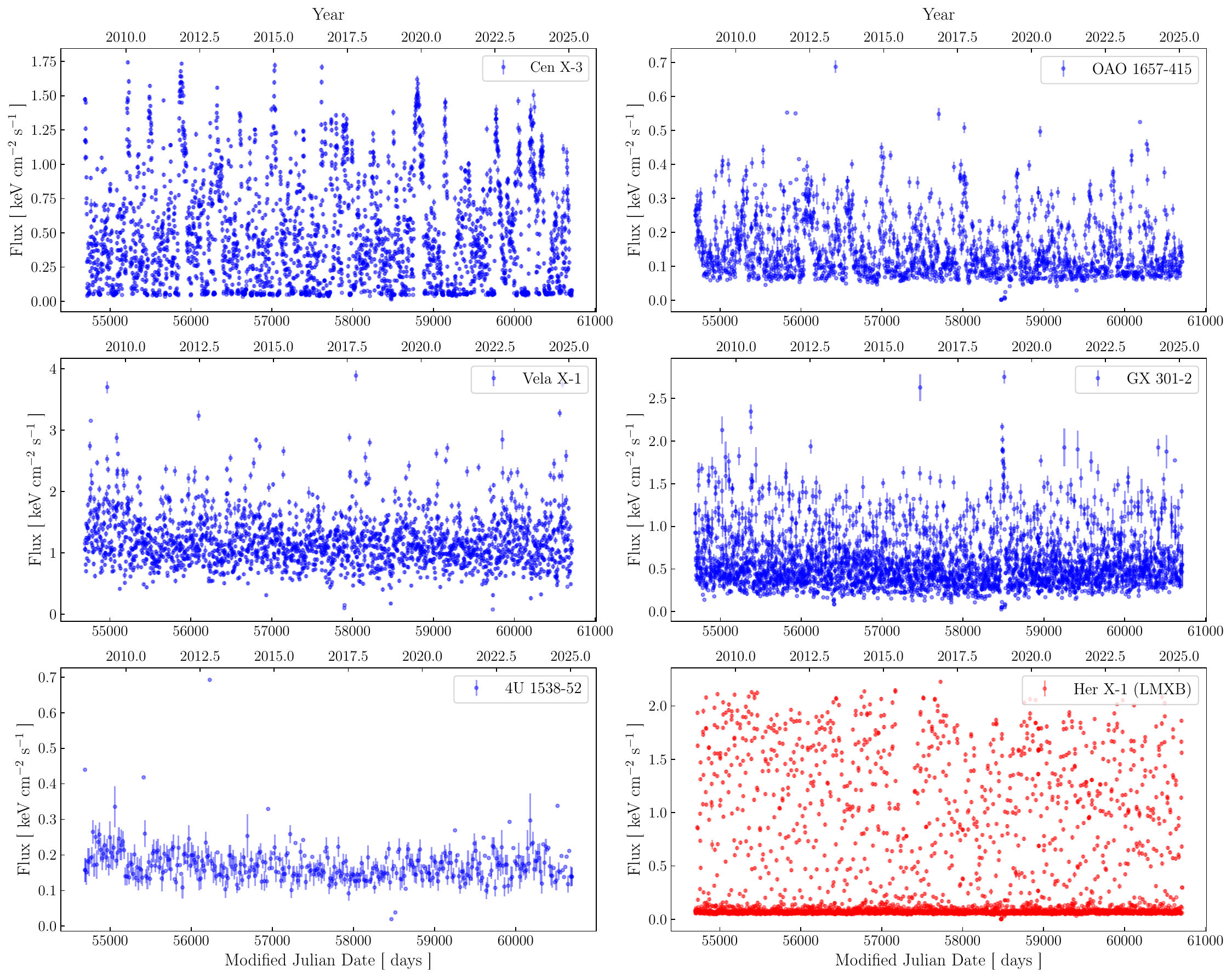}}
\caption{Flux from some HMXB persistent pulsars (in keV cm$^{-2}$ s$^{-1}$) for a time in Modified Julian Date (MJD) observed by Fermi-GBM in the range $12$--$50$~keV (Fermi Collaboration 2022). The persistent pulsars, unlike transients, do not demonstrate a strict correlation between their X-ray flux and changes in spin rotation \citep{Bildsten...1997ApJS..113..367B}.  The case of LMXB pulsar Her~X-1 is further discussed in the Section~\ref{sub_LMXB}. }
\label{fig:hmxb_flux}
\end{figure}

The article is divided into several sections. Section~\ref{sec:spin} discusses spin periods and their changes observed in accreting X-ray pulsars in HMXB systems from the Galactic population. Section~\ref{sec:acc} is devoted to considering various accretion regimes and possible accretion structures surrounding X-ray pulsars in massive binary systems. Theoretical estimates of the torques transferred by the accretion flow to the NS providing its spin evolution are presented. Section~\ref{sec:orb} covers the range of orbital periods for X-ray pulsars in HMXB systems. The focus is on persistent sources and orbital decay, and its implications for these systems are discussed. Section~\ref{sec:select} highlights five persistent X-ray pulsars from the Galactic HMXB population, we selected objects (Fig.~\ref{fig:image1}--\ref{fig:hmxb_flux}) with the longest history of observations of their spin periods. One of them is RLO pulsar (Cen~X-3) and four are sgXRB pulsars (OAO~1657$-$415, Vela~X-1, 4U~1538$-$52 and GX~301$-$2). Section~\ref{sec:disc} revolves around the spin evolution of NSs in HMXB systems and comparison with X-ray pulsars in LMXB systems (using Her~X-1 as an example). It synthesizes findings from previous sections and delves into how NS spin rates change over time due to their interaction with massive companions. The Conclusion in Section~\ref{sec:conc} summarises the complex processes driving spin evolution in X-ray pulsars. It emphasizes the dynamic interaction between the NS and the massive companion star, which influences the star's spin behavior. The results presented in this paper may be useful for studies in the field of evolution of massive binary systems with a degenerate companion, as well as for studying accretion processes and their influence on the spin evolution of X-ray pulsars.


\section{Pulsar spin periods and their changes} \label{sec:spin}

Spin periods of accreting X-ray pulsars in the Galatic population of HMXB systems cover an extensive range of values from $2.76$ to $36200$~s, excluding SAX~J0635.2$+$0533, whose X-ray radiation has a non-accretionary nature \citep{Mereghetti...2009AA...504..181M}. The values of spin periods of these sources do not remain constant but undergo regular changes (variations) due to the interaction between an NS and accretion flow, leading to spin-up and spin-down processes in axial (spin) rotation. This phenomenon is mainly caused by the process of torque transfer ($\overline{K} = \overline{r}\times\overline{F}$) from the accretion flow to the NS. In particular, this leads to a change in the rotation frequency of the neutron star $\nu = 1/P_{s}$, in accordance with the expression \citep{Kornilov...1983SvA....27..163K}:
\begin{equation}\label{1.4.1.1} 
\left| K \right| = I\,\dot{\omega}_{\rm s} = 2 \pi\, I\,\left| \dot{\nu_{\rm s}}\right| ,
\end{equation} 
where $I$ is the moment of inertia of the NS, $\dot{\nu} = d\nu/dt$ is the rate of change of the NS spin frequency (spin period). The efficiency of torque transfer from the accretion flow to the NS depends significantly on the implementation of specific accretion approximations, which were considered in the previous section.

\subsection{Local and global spin-up and spin-down trends} \label{sec:loc_glob}

Local spin-up and spin-down trends are episodic, chaotic variations in the NS spin period. They occur at a relatively high rate, typically an order of magnitude larger than the rates associated with longer-term trends, and last from several days to a few months. Meanwhile, global spin-up and spin-down trends span months to years, generally appearing as a slower overall drift in the spin period. 

The fact that local trends occur against the background of global trends—and are larger in magnitude—suggests that the NS operates near an equilibrium spin period, where spin-up and spin-down torques largely balance \citep{Ikhsanov...2014ARep...58..376I}. Non-periodic fluctuations in the accretion rate, driven by inhomogeneities in the accretion flow, give rise to these short-term changes in spin \citep{Bozzo...2016AA...589A.102B}.

However, the equilibrium period itself may also evolve on longer timescales, resulting in gradual spin-up or spin-down trends. Such long-term drifts could be linked to systematic variations in the stellar wind from the massive companion \citep{Chandra...2021MNRAS.508.4429C}. Because the accretion torque in wind-fed systems depends strongly on the wind velocity (e.g., $\dot{\nu}_{\rm s}\propto K\propto \upsilon_{\rm rel}^{-4}$), even small changes in that velocity can produce noticeable shifts in the spin rate (see Section~\ref{sec:acc}, Eq.~\ref{sa_qsp}--\ref{nu_qsp_up}). Similar multi-year modulations in flow velocity are also observed in the solar wind, tied to its 11-year activity cycle \citep{Li...2017MNRAS.472..289L}.

\subsection{Super-global spin-up trends} \label{sec:superglob}

In the 1980s, it was noted that many known HMXB X-ray pulsars tend to spin up \citep{Lipunov...1992ans..book.....L}. As more data were collected, this general pulsar spin-up became increasingly apparent, extending over timescales of decades and exceeding the multi-year durations of typical ``global'' spin-up trends. We introduce the term a \emph{super-global spin-up trend} to describe this phenomenon. Previous studies have proposed several hypotheses regarding its origin:

\begin{itemize} 
    \item \textbf{Stellar evolution of the massive component.} \citet{Siuniaev...1977SvAL....3..114S} suggested that the stellar wind outflow from the massive companion could intensify as the star evolves, thereby accelerating the NS. However, \citet{Lipunov...1992ans..book.....L} criticized this idea, arguing that stellar evolution typically has minimal impact on a timescale of just a few decades, making this explanation unlikely.
    \item \textbf{Selection effects.} In scenarios where a Keplerian accretion disk is formed, spin acceleration occurs while the NS is in a bright accretion phase \citep{Ghosh...1979ApJ...232..259G, Ghosh...1979ApJ...234..296G}, whereas spin deceleration takes place when the source becomes faint (e.g., in a propeller regime). However, since most accreting X-ray pulsars rotate near an equilibrium period---often longer than the threshold needed to trigger the propeller state---this explanation also seems dubious.
    \item \textbf{Asymmetric spin-up and spin-down phases.} The spin-up phase may be relatively prolonged but gradual, whereas the spin-down phase is shorter and more rapid \citep{Lipunov...1987ApSS.132....1L}. Consequently, one is more likely to observe the pulsar during spin-up. This disparity can arise if the scalar potential on the accretion-flow side is asymmetric, regardless of the accretion regime \citep{Lipunov...1992ans..book.....L}. Within the framework of this approach, on large time scales, there should be an alternation of protracted spin-up trends with relatively short but fast rotation deceleration trends, while the average value of the equilibrium period should remain the same (the general trend of the period change should be near zero). However, observations of the overwhelming majority of sources under consideration indicate a general trend towards spin-up.
\end{itemize}

\begin{figure}[ht]
\center{\includegraphics[width=\textwidth]{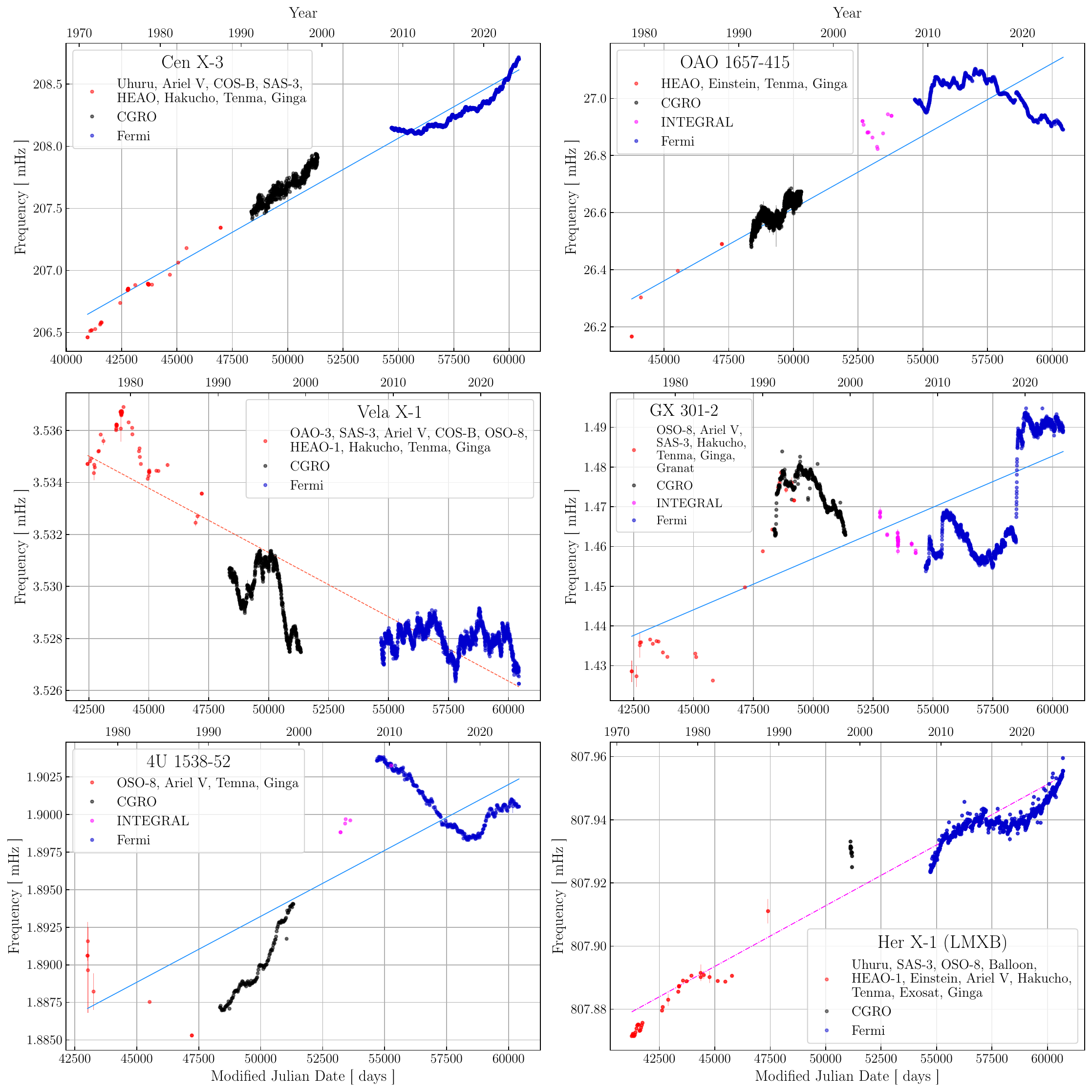}}
\caption{Evolution of the spin frequency of long-observed Galactic X-ray pulsars in HMXBs and LMXB pulsar Her~X-1. The solid blue line corresponds to the super-global trend, according to values shown in Table~\ref{tab:pulsar_nu_dot_super}, obtained by the KDE-weighted regression model in linear approximation. A negative trend of spin frequency by Vela~X-1 (dashed red) is discussed in the text. A unique case of long-monitored LMXB system Her~X-1 and its trend (dot-dashed magenta) is shown on the lower right panel. Data was collected within the time range from 1970s to 2024 received by HEAO-1, Tenma, Ginga, Einstein, OSO-8, Ariel~V, and Granat \citep[red;][]{Nagase...1989PasJ...396..P147}, the CGRO \citep[black;][]{Batse}, INTEGRAL \citep[magenta;][]{Integral_pulsars}, and Fermi-GBM \citep[blue;][]{Fermi_pulsars}. Measurement uncertainties for data of newer generations of space telescopes stay within the data points.}
\label{fig:image1}
\end{figure}

Fig.~\ref{fig:image1} illustrates examples of the spin evolution of X-ray pulsars in HMXBs with long observational histories. It reflects time-sparse observations by pioneering missions on the dawn of X-ray astrophysics compared to time-dense observations and data by currently operating space observatories with improved electronics onboard. Therefore, to fit non-uniformly distributed and varying density data, we have used a \emph{KDE-weighted} (Kernel Density Estimator) regression model. We have performed a three-step procedure with KDE to estimate density at each data point, then assigning weights that are inversely proportional to the data densities, and fitting the data with a weighted regression model using these defined weights. For the KDE itself, we employed a Gaussian kernel with bandwidths between 200 and 2000 days, optimized via cross-validation\footnote{The initial bandwidth was chosen between 1\% and 10\% of the overall observational span ($\sim$20{,}000 days).}.

The resulting fits, the solid blue lines in Fig.~\ref{fig:image1}, capture the super-global spin-up trends (or spin-down in Vela~X-1, shown by a dashed red line). Converting these slopes to frequency derivatives demonstrates the change in neutron-star spin parameters. Table~\ref{tab:pulsar_nu_dot_super} reports the super-global rates over corresponding Modified Julian Date (MJD) intervals. Notably, all pulsars in the sample (except Vela~X-1) show a net spin-up on decadal or longer timescales.

In this paper, we propose a new hypothesis to explain the phenomenon of super-global spin-up trends of X-ray pulsars in HMXB systems. There, the observed behavior is interpreted to be a result of orbital angular momentum transfer to the NS’s spin angular momentum via tidal interactions with its massive companion and accretion. In the next Section, we consider possible accretion regimes and structures, as well as the efficiency of the torques applied by the accretion flow to the NS, affecting its spin evolution.

\begin{deluxetable}{lccl}
\tabletypesize{\scriptsize}
\tablewidth{0pt} 
\tablecaption{Estimates of the super-global spin frequency change $\dot{\nu}_{\rm s}$ of X-ray pulsars in HMXBs and LMXB pulsar Her~X-1 for corresponding timescales in Modified Julian Date. \label{tab:pulsar_nu_dot_super}}
\tablehead{
    \colhead{Object} & \colhead{$\dot{\nu}_{\rm s}$} & \colhead{MJD} & \colhead{References}\\
\colhead{}& \colhead{(Hz s$^{-1}$)} & \colhead{(days)} & \colhead{}}
\startdata 
        Cen~X-3   & $1.16$E-12  & $40960$--$60441$  & a, b, c \\
        OAO~1657$-$415  & $5.10$E-13  & $43756$--$60419$  & a, b, d, e\\
        \textbf{Vela~X-1}  & $-5.60$E-15  &  $42499$--$60438$ &  a, b, f\\
        4U~1538$-$52   & $1.18$E-14 &  $43016$--$60420$ &  a, b, g \\
        GX~301$-$2  & $3.00$E-14  & $42417$--$60436$ & a, b, d, h\\
        \hline
        Her~X-1  & $4.39$E-14  & $41257$--$60707$ & a, b, i\\
      \enddata
      \tablecomments{[a] \cite{Nagase...1989PasJ...396..P147}, [b] \cite{Batse}, [c] \cite{Fermi_GBM_Cen_X_3}, [d] \cite{Integral_pulsars}, [e] \cite{Fermi_GBM_OAO-1657}, [f] \cite{Fermi_GBM_Vela_X_1}, [g] \cite{Fermi_GBM_4U_1538_52}, [h] \cite{Fermi_GBM_GX_301_2}, [i] \cite{Fermi_GBM_Her_X_1}.}
\end{deluxetable}


\section{Accretion Regimes, Accretion Structures and Their Predicted Torques}
\label{sec:acc}
\subsection{Accretion Regimes}
\label{sub:regimes}
Accretion is the primary energy source in X-ray binaries and it plays a central role in determining NSs' X-ray emission and spin evolution \citep{Frank...2002apa..book.....F}. The type of accretion regime depends on several factors, including the nature of the massive component, the orbital parameters, and the NS's parameters: magnetic field, spin period, etc. \citep{Karino...2019PASJ...71...58K}. This section discusses the key accretion mechanisms for HMXB pulsars and their implications for NS spin evolution.
\begin{itemize}
    \item \textit{Roche-Lobe Overflow (RLO) Accretion}\\
    In this scenario, mass transfer occurs when the donor star fills its Roche lobe at a 
late stage of its evolution. Matter is lost from the star at a high rate, flowing 
through the inner Lagrange point (L1) into the Roche lobe of the compact object. This transferred matter can form either a channeled stream or an accretion disk around the compact object \citep{Fornasini...2023hxga.book..143F, Blondin...1997ASPC..121..361B}.  
Crucially, in the RLO accretion regime, the outflowing matter possesses significant 
angular momentum because the radius of gravitational capture is of the order of 
the semimajor axis ($r_{\rm _G} \approx a$).

The RLO accretion regime is primarily observed in low-mass X-ray binary (LMXB) systems, where the compact object is typically an NS or a white dwarf. The donor star is a late-type (for most cases, K/M spectral type) star located on or near the Main Sequence \citep{Fortin...2024AA...684A.124F}.
In some cases, this accretion regime can also occur in HMXB pulsar systems, such as Cen X-3, SMC X-1 and LMC X-4. In these systems, the massive donor star fills its Roche lobe, leading to the formation of a steady accretion disk around the NS \citep{Fornasini...2023hxga.book..143F}.

\item\textit{Be X-Ray Binary Mechanism}\\
In Be X-ray binaries, the donor star (massive component) is a rapidly rotating, massive, early-type star (B spectral class) that exhibits Balmer emission lines in its spectrum. This star is surrounded by a decretion disk, which forms as a result of material being ejected due to a combination of different factors: rapid rotation, stellar wind and non-radial pulsations \citep{Rast...2025MNRAS.537.3575R, Semaan...2018AA...613A..70S}. The NS in a Be X-ray binary interacts with the decretion disk of its massive companion, particularly during periastron passages, leading to a significant increase in the accretion rate \citep{Reig...2011Ap&SS.332....1R, Okazaki...2013PASJ...65...41O}. This phenomenon is characteristic of systems with large orbital eccentricities ($e>0.3$) and is commonly observed in transient X-ray pulsars, which display episodic outburst activity.
However, there are exceptions \citep{Reig...2011Ap&SS.332....1R} where Be X-ray binaries host persistent X-ray pulsars, exhibiting relatively stable X-ray flux over time despite the presence of a decretion disk of a massive companion. These systems demonstrate the diverse accretion behaviors possible in Be X-ray binaries. The observed characteristics of transient pulsars can be well interpreted in the framework of quasi-spherical accretion and, in some cases, in the framework of standard disk accretion \citep{Reig...2011Ap&SS.332....1R} and accretion from MAD-disk \citep{Ikhsanov...2015MNRAS.454.3760I}. As the NS moves along its orbit, it can also capture matter from the stellar wind of its massive companion; thus, the Be X-Ray Binary mechanism is complementary to the wind accretion regime.

\item\textit{Wind Accretion}\\
If the massive donor does not fill its Roche lobe, accretion proceeds from its radiatively driven wind. Early-type O/B supergiants typically lose $\sim 10^{-5}$--$10^{-7}\, M_{\odot}\, \text{yr}^{-1}$ \citep{Martinez...2017SSRv..212...59M}, while luminous O giants and the most massive O~V stars reach $\sim 10^{-7}$--$10^{-8}\, M_{\odot}\, \text{yr}^{-1}$ \citep{Ribo...2006AA...449..687R} and Be/B~V stars are usually an order of magnitude lower \citep{Bernacca...1981AA....94..345B}. As the NS orbits its companion, it captures some of the matter from the stellar wind and accretes it onto its surface. This mass transfer mechanism is known as wind accretion \citep{Davidson...1973ApJ...179..585D} and is the dominant mechanism for most X-ray pulsars in HMXB systems.
A variety of parameters can influence how the accretion flow evolves, but wind accretion regime is typically described by one of three main approximations: quasi-spherical accretion, a Keplerian (standard) disk, or a magnetically arrested (MAD) disk. These models are based on the fundamental conservation laws of mass, angular momentum, and magnetic flux. If we ignore the accretion flow’s own magnetic field, while still accounting for the angular momentum of the captured matter, then the scenario reduces to the traditional non-magnetic approximations: either quasi-spherical accretion \citep{Elsner...1977ApJ...215..897E, Shakura...2015ARep...59..645S} or accretion through a Keplerian disk \citep{Shakura...1973SvA....16..756S}.
\end{itemize}

\subsection{Possible accretion structures and their interaction with a magnetized neutron star}
\subsubsection{Quasi-spherical accretion} \label{QSP}
In the absence of angular momentum and a magnetic field in the accretion flow, the flow can be described using the spherically symmetric approximation. This model was proposed in \citet{Bondi...1952MNRAS.112..195B}, predating the discovery of X-ray pulsars. Under such conditions, matter moves radially and uniformly from all directions toward the star in a free-fall regime \citep{Zeldovich...1969SvA....13..175Z}. However, this approximation is highly idealized and not applicable to binary systems, where the accreting matter inherently possesses angular momentum due to the orbital motion of the components. The modification of spherically symmetric accretion to account for the angular momentum of the flow can be described within the framework of the quasi-spherical accretion approximation \citep[QSp,][]{Shakura...2013PhyU...56..321S}.  According to \citet{Shakura...2015ARep...59..645S}, there are two scenarios of QSp: Bondi-Hoyle-Littleton (BHL) accretion (or supersonic accretion) and subsonic accretion. The supersonic scenario occurs when the accreting matter cools rapidly mainly due to Compton processes and moves in a free-fall regime towards the NS magnetosphere. The velocity of the matter quickly exceeds the sound speed ($\upsilon_{\rm ff} > c_{\rm s}$), forming a shock wave at some distance above the magnetosphere. Subsonic accretion occurs if the captured matter does not have time to cool rapidly, so that: $\upsilon_{\rm ff} < c_{\rm s}$. In this case, a hot quasi-spherical shell is formed around the NS magnetosphere.

According to \citet{Illarionov...1975A&A....39..185I}, angular momentum captured by an NS per unit time (torque) from its massive companion's stellar wind on the Bondi radius regardless of the realized accretion structure  corresponds to:
\begin{equation}\label{sa_qsp}
 K^{\rm (w)} = \dot{M} \, \Omega_{\rm orb}\, r_{\rm_G}^2
\end{equation}
Here, $r_{\rm _G}=2GM_{\rm ns}/\upsilon_{\rm rel}^2$ is the Bondi radius, where $G$ is a gravitational constant, $M_{\rm ns}$ is the mass of the NS, $\upsilon_{\rm rel} = \sqrt{\upsilon_{\rm w}^2 + \upsilon_{\rm orb}^2}$ is relative velocity of a NS as a vector sum of wind velocity $\upsilon_{\rm w}$ and NS orbital velocity $\upsilon_{\rm orb}$, $\dot{M}$ is a rate of gravitational capture of matter by a NS from the stellar wind, $ \Omega_{\rm orb} = 2\pi/P_{\rm orb}$ is an orbital angular velocity with $P_{\rm orb}$ orbital period. 

Solving Eq.~\ref{sa_qsp} concerning Eq.~\ref{1.4.1.1} allows us to estimate the maximal possible (in absolute value) spin-up and spin-down rates of an NS accreting from stellar wind in HMXB system:

\begin{eqnarray}\label{nu_qsp_up}
 \dot{\nu}^{\rm (w)}  &=& \frac{\left|K^{\rm (w)} \right|}{2\pi I} \simeq 1 \times 10^{-11} \, \text{Hz s}^{-1}\, \times\, m^{2}\, I_{45}^{-1}\, \times \nonumber\\
& \times & \left(\frac{P_{\rm orb}}{10.45\,\text{d}}\right)^{-1}
\left(\frac{\dot{M}}{3\times10^{16}\,\text{g s}^{-1}}\right) \left(\frac{\upsilon_{\rm
rel}}{270\,\text{km s}^{-1}}\right)^{-4}
\end{eqnarray}

In Eq.~\ref{nu_qsp_up} and further equations, for clarity, we used the parameters of the well-known X-ray pulsar OAO~1657$-$415 (see Table \ref{tab:pulsar_opt} and Table~\ref{tab:pulsar_mass_comp}). Here, $I_{45} = I/10^{45}\, \text{g cm}^2$ is normalized moment of inertia of NS, $m = M_{\rm ns}/1.4 M_{\odot}$ is the mass of NS normalized to canonical value, $\dot{M} = L_{\rm x} R_{\rm ns}/GM_{\rm ns}$ is accretion rate, $\mu_{\rm ns} = 0.5 B_{\rm ns} R_{\rm ns}^3$ is NS dipole magnetic moment. From Eq.~\ref{nu_qsp_up}, it is clear that the value of the upper limit of spin changes within the framework of the wind accretion regime depends significantly on the wind velocity of the massive companion.

As the accreting matter moves from the Bondi radius $r_{\rm _G}$ to the boundary of the NS magnetosphere with radius $r_{\rm m}$, partial dissipation of the angular momentum occurs. In the case where the flow retains its quasi-spherical structure in the interval $r_{\rm _G} > r > r_{\rm m}$ (without transforming into a disk), the torque applied from the quasi-spherical flow to the boundary of the magnetosphere corresponds to the expression \citep{Ruffert...1999A&A...346..861R, Ikhsanov...2014ARep...58..376I}:

\begin{equation}\label{qsp_k}
K^{\rm (qsp)}  =  \xi \, K^{\rm (w)} 
\end{equation}
Here, $\xi$ is a dimensionless parameter characterizing the dissipation of angular momentum in a quasi-spherical accretion flow.

In the case of the BHL accretion, according to \citet{Shakura...2013PhyU...56..321S}, depending on the direction of the specific angular momentum vector of the captured matter (in the direction or against the direction of the orbital angular momentum), the rotation of the NS can spin up or spin down. In the case of subsonic accretion, when a quasi-spherical shell is formed, turbulent motions of matter interacting with the NS magnetosphere can cause both spin-up and spin-down of the NS's rotation even if the angular momentum of the captured matter is co-directed with the orbital angular momentum \citep{Shakura...2013PhyU...56..321S}. The maximum (in absolute value) possible spin-down torque applied to NS magnetospheric boundary in this scenario is: 

\begin{equation}\label{sd_qsp}
 K_{\rm sd}^{\rm (qsp)}  = - \dot{M} \,\omega_{\rm
s}\,r_{\rm_A}^2
\end{equation}
Here, $r_{\rm_A} = (\mu_{\rm ns}^{2}/ \dot{M}\, \sqrt{2GM_{\rm
ns}})^{2/7}$ is the Alfven radius, which in classical accretion scenarios is taken as the radius of the magnetosphere $r_{\rm m} \sim r_{\rm _A}$ \citep{Lipunov...1992ans..book.....L, Frank...2002apa..book.....F}. 

Solving Eq.~\ref{qsp_k} and Eq.~\ref{sd_qsp} concerning Eq.~\ref{1.4.1.1}, allows us to estimate the upper limits of spin-up and spin-down rates of the NS within the framework of QSp accretion scenario:

\begin{eqnarray}\label{nu_qsp_up2}
 \dot{\nu}_{\rm su}^{\rm (qsp)}  &=&  \xi \, \dot{\nu}^{\rm (w)}_{\rm su} \simeq 2 \times 10^{-12} \, \text{Hz s}^{-1}\, \times\, \xi_{0.2} \, m^{2}\, I_{45}^{-1}\, \times \nonumber\\
& \times & \left(\frac{P_{\rm orb}}{10.45\,\text{d}}\right)^{-1}
\left(\frac{\dot{M}}{3\times10^{16}\,\text{g s}^{-1}}\right) \left(\frac{\upsilon_{\rm
rel}}{270\,\text{km s}^{-1}}\right)^{-4}
\end{eqnarray}

\begin{eqnarray}\label{nu_qsp_down}
 \dot{\nu}_{\rm sd}^{\rm (qsp)}  &=& \frac{K^{\rm (qsp)}_{\rm sd}}{2\pi I} \simeq - 3 \times 10^{-13} \, \text{Hz s}^{-1}\, \times\, m^{-2/7}\, I_{45}^{-1}\, \times \nonumber\\
& \times & \left(\frac{P_{\rm s}}{37\,\text{s}}\right)^{-1}
\left(\frac{\dot{M}}{3\times10^{16}\,\text{g s}^{-1}}\right)^{3/7} \left(\frac{\mu_{\rm
ns}}{1.6\times10^{30}\,\text{G cm}^3}\right)^{8/7}
\end{eqnarray}
Here, $\xi_{0.2} = \xi/0.2$ is a normalised parameter of angular momentum dissipation in a quasi-spherical accretion flow estimated in \citep{Ruffert...1999A&A...346..861R} by means of numerical simulations.

In the accretion state, regardless of the realized accretion scenario, the NS evolves over time to a spin period called the equilibrium period $P_{\rm eq}$ \citep{Lipunov...1992ans..book.....L, Shakura...2012MNRAS.420..216S}. This period is reached when the sum of torques applied to the NS becomes zero:

\begin{equation}\label{eq_p}
I\,\dot{\omega} = \sum K \equiv 0
\end{equation}

However, even minor fluctuations in the velocity and density of the captured stellar wind will significantly affect the balance of the total torque in Eq.~\ref{eq_p}, thereby provoking episodic variations of the spin period changes. According to \citet{Shakura...2013PhyU...56..321S}, a value of the equilibrium period of the NS in the QSp accretion scenario also depends significantly on the stellar wind velocity captured by the NS.

\subsubsection{Accretion from Keplerian (standard) disk}
\label{sect_kd}
Keplerian disk (Kd) accretion structure can form in systems where a massive companion fills its Roche lobe (with RLO mechanism) \citep{Fornasini...2023hxga.book..143F}, as well as in systems with Be X-Ray Binary Mechanism \citep{Okazaki...2013PASJ...65...41O} and in systems with wind accretion \citep{Baykal...1997A&A...319..515B, Karino...2019PASJ...71...58K, Shakura...2013PhyU...56..321S, Ikhsanov...2013ARep...57..287I}.

In the Kd, excess angular momentum interferes with the free radial movement of matter. Its further movement occurs as the angular momentum dissipates \citep{Shakura...1973SvA....16..756S, Pringle...1972A&A....21....1P}. If the formation of the Keplerian disk occurs in the wind accretion regime, then the angular momentum captured by an NS per unit time (torque) corresponds to Eq.~\ref{sa_qsp}. In the case of the RLO accretion, when the massive companion fills its Roche lobe, flowing out through the inner Lagrange point (L1), the matter has a significantly greater angular momentum (compared to wind accretion). In this case, the radius of matter capture is comparable to the radius of the semimajor axis ($r_{\rm_G} \approx a$), the torque transferred by the matter corresponds to the expression \citep{Gorbatskii...1977SvA....21..587G, Lipunov...1992ans..book.....L}:

\begin{equation}\label{K_RLO}
  K^{\rm (rlo)} \sim \dot{M}\, \Omega_{\rm orb}\, a^2      
\end{equation}

Solving Eq.~\ref{K_RLO} with the concerning Eq.~\ref{1.4.1.1} allows us to estimate the upper limit of spin-up rate of the NS within the framework of RLO accretion regime:

\begin{eqnarray}\label{nu_RLO}
     \dot{\nu}^{\rm (rlo)}  &=& \frac{\left|K^{\rm (rlo)}\right|}{2\pi I} \sim  4\times 10^{-10} \, \text{Hz  s}^{-1} \times \, I_{45}^{-1}\, \left(\frac{\dot{M}}{3\times10^{16}\,\text{g s}^{-1}}\right) \, \times \nonumber\\
& \times & \left(\frac{P_{\rm orb}}{10.45\,\text{d}}\right)^{1/3} \left(\frac{M_{\rm ns} + M_{\rm opt}}{15.7\,\odot}\right)^{2/3}
\end{eqnarray}

The maximal (in absolute value) possible spin-down torque applied to NS from accretion flow on the boundary of the magnetosphere in the case of Kd is \citep{Lynden...1974MNRAS.168..603L}: 

\begin{equation}\label{sd_kd}
K_{\rm sd}^{\rm (kd)}  = - k_{\rm
t} \frac{\mu_{\rm ns}^2}{r_{\rm cor}^3}
\end{equation}
here, $r_{\rm cor} = (2GM_{\rm ns}/\omega_{\rm s}^2)^{1/3}$ is corotation radius of an NS.
The maximal possible spin-up torque applied to NS on the magnetospheric boundary \citep{Pringle...1972A&A....21....1P}:

\begin{equation}\label{su_kd}
K_{\rm su}^{\rm (kd)}  = \dot{M}\sqrt{GM_{\rm ns}\, r_{\rm_A}}
\end{equation}

Solving Eq.~\ref{sd_kd} and Eq.~\ref{su_kd} concerning Eq.~\ref{1.4.1.1} allows us to estimate the upper limits of spin-up and spin-down rates of the NS within the framework of Kd accretion scenario:

\begin{equation}\label{nu_kd_up}
 \dot{\nu}_{\rm su}^{\rm (kd)}  = \frac{K^{\rm (kd)}_{\rm su}}{2\pi I} \simeq 2 \times 10^{-12} \, \text{Hz s}^{-1}\, \times\, m^{3/7}\, I_{45}^{-1}\,
\left(\frac{\dot{M}}{3\times10^{16}\,\text{g s}^{-1}}\right)^{6/7} \left(\frac{\mu_{\rm ns}}{1.6\times10^{30}\,\text{G cm}^3}\right)^{2/7}
\end{equation}

\begin{equation}\label{nu_kd_down}
     \dot{\nu}_{\rm sd}^{\rm (kd)}  = \frac{K^{\rm (kd)}_{\rm sd}}{2\pi I} \simeq - 6 \times 10^{-14} \, \text{Hz s}^{-1}\, \times\,k_{\rm t}\, m^{-1}\, I_{45}^{-1} \left(\frac{P_{\rm s}}{37\,\text{s}}\right)^{-2}  \left(\frac{\mu_{\rm ns}}{1.6\times10^{30}\,\text{G cm}^3}\right)^{2}
\end{equation}

Solving Eq.~\ref{eq_p} using Eq.~\ref{sd_kd} and Eq.~\ref{su_kd} gives the following estimate of the equilibrium period in Kd scenario:

\begin{equation} \label{kd_eq}
P_{\rm eq}^{\rm (kd)} = 2\pi \, k_{\rm t}^{1/2} \left[\frac{2^{1/4}\, \mu_{\rm ns}^{6}}{(GM_{\rm ns})^{5} \, \dot{M}^{3}} \right]^{1/7}
\end{equation}

The equilibrium period of the NS, according to Eq.~\ref{kd_eq}, does not exceed several tens of seconds for the canonical parameters of an NS and a magnetic field of $10^{12}$ G. However, most of the observed HMXB pulsars have significantly larger spin periods, reaching hundreds and even thousands of seconds. In \citet{Ruffert...1997A&A...317..793R, Shakura...2013PhyU...56..321S}, the possibilities of forming retrograde Keplerian disks are considered, which can form due to inhomogeneities in the stellar wind and lead to the rotation of pulsars to large values of spin periods. However, the existence of retrograde disks in such systems is a subject of discussion \citep{Shakura...2013PhyU...56..321S}.

\subsubsection{Accretion from Magnetically-Arrested (MAD) Disk}
The accretion structure may differ significantly from classical ones if the matter captured from the stellar wind at the Bondi radius has its magnetic field \citep{ Ikhsanov...2025AstBu..80..122I, Shvartsman...1971SvA....15..377S, Bisnovatyi-Kogan...1974Ap&SS..28...45B, Bisnovatyi...2019Univ....5..146B}. 
The accretion from the Magnetically-Arrested (MAD) Disk (also known as Magnetic Levitational Disk or ML-disk) is realized under the condition that the magnetic field strength in the flow is high enough \citep{Ikhsanov...2015MNRAS.454.3760I}. As was shown in \citet{Bisnovatyi...1976Ap&SS..42..401B}, in this case, a non-Keplerian Disk can form, in which the motion of the matter in the radial direction is impeded by its proper magnetic field.

In the case of accretion from MAD-disk, the torque applied to NS from accretion flow on NS magnetospheric boundary is \citep{Ikhsanov...2015MNRAS.454.3760I}: 

\begin{equation}\label{k_total}
K^{(\rm ml)} = k_{\rm t} \frac{\mu_{\rm ns}^2}{(r_{\rm cor}\, r_{\rm
ma})^{3/2}} \left[ \frac{\Omega_{\rm f}(r_{\rm ma})}{\omega_{\rm s}}
- 1 \right]
\end{equation}
Here, $r_{\rm ma}$ is the inner radius of MAD-disk, $\Omega_{\rm f}(r_{\rm ma})$ is the angular velocity of accreting matter on the $r_{\rm ma}$ radius. 

Depending on the ratio $[\Omega_{\rm f}(r_{\rm m})/\omega_{\rm s}]$ the torque from Eq.~\ref{k_total} can be either spin-up (if angular velocity of accretion flow faster than spin angular velocity) or spin-down (in the opposite situation: $\Omega_{\rm f}(r_{\rm m})<\omega_{\rm s}$). In case the NS spin angular velocity significantly exceeds the angular velocity of accretion flow $\omega_{\rm s} >> \Omega_{\rm f}(r_{\rm m})$ the Eq.~\ref{k_total} transforms to maximal possible spin-down torque \citep{Ikhsanov...2012ApJ...753....1I}:

\begin{equation}\label{2.2.54}
K_{\rm sd}^{(\rm ml)} = - k_{\rm t} \frac{\mu_{\rm
ns}^2}{(r_{\rm cor}\, r_{\rm ma})^{3/2}}
\end{equation}

Since we consider accretion from the MAD-disk in the wind accretion regime, where the angular momentum captured from the stellar wind corresponds to Eq.~\ref{sa_qsp} (see Section~\ref{QSP}), the maximum possible spin-up rate will also correspond to Eq.~\ref{nu_qsp_up}. The maximum possible spin-down rate within the framework of this model can be estimated from the joint solution of Eq.~\ref{1.4.1.1} and Eq.~\ref{2.2.54}:

\begin{eqnarray}\label{4.7}
\dot{\nu}_{\rm sd}^{\rm (ml)}  & = & \frac{K^{\rm (ml)}_{\rm sd}}{2\pi I} \simeq - 1 \times 10^{-11}\,\text{Hz s}^{-1} \,\times\,k_{\rm t}\, m^{-8/13} \, T_{6}^{3/13} \, \alpha_{0.1}^{-3/13} \times \nonumber\\
& \times &  \left(\frac{P_{\rm s}}{37\,\text{s}}\right)^{-1} \left(\frac{\dot{M}}{3 \times 10^{16}\,\text{g s}^{-1}}\right)^{6/13} \left(\frac{\mu_{\rm ns}}{1.6 \times 10^{30}\,\text{G cm}^3}\right)^{17/13}
\end{eqnarray}

The formulas for estimating the equilibrium period of pulsars in the MAD-disk accretion scenario are quite cumbersome. Their derivation and subsequent analysis are described in detail in \citet{Ikhsanov...2015MNRAS.454.3760I}. The same work shows that in the MAD-disk accretion mode, the transfer of angular momentum from the accretion flow to the NS can be more efficient, thereby explaining the rapid spin evolution of some HMXB pulsars.

\subsection{Some conclusions and remarks}

The amount of angular momentum captured with the matter by an NS at the radius of matter capture depends on the accretion regime. In the case of wind accretion, regardless of the accretion structure formed around the NS, the torque does not exceed the value specified in Eq.~\ref{sa_qsp}. Thus, the maximum possible spin-up rate of the NS within the wind accretion corresponds to Eq.~\ref{nu_qsp_up}. In the case of the RLO accretion mode with the formation of a Keplerian disk, the torque transferred to the NS does not exceed the value specified in Eq.~\ref{K_RLO}, and the maximum possible spin-up rate corresponds to Eq.~\ref{nu_RLO}. The efficiency of angular momentum transfer from the captured matter to the NS depends on the formed accretion structure. According to conservation laws, the angular momentum transferred to the NS by the accretion flow cannot exceed the angular momentum captured at the Bondi radius. Thus, the observed rates of spin period changes should not exceed theoretical limits within the framework of the realized accretion scenario. Namely:

\begin{itemize}
    \item Wind accretion with QSp structure: $|K^{\rm (w)}| \geq |K^{\rm (qsp)}| \geq |K^{\rm (obs)}| \, \Rightarrow \, |\dot{\nu}^{\rm (w)}| \geq |\dot{\nu}^{\rm (qsp)}| \geq |\dot{\nu}^{\rm (obs)}|$ 
    \item Wind accretion with Kd structure: $|K^{\rm (w)}| \geq |K^{\rm (kd)}| \geq |K^{\rm (obs)}| \, \Rightarrow \, |\dot{\nu}^{\rm (w)}| \geq |\dot{\nu}^{\rm (kd)}| \geq |\dot{\nu}^{\rm (obs)}|$ 
    \item Wind accretion with MAD-disk structure: $|K^{\rm (w)}| \geq |K^{\rm (ml)}| \geq |K^{\rm (obs)}| \, \Rightarrow \, |\dot{\nu}^{\rm (w)}| \geq |\dot{\nu}^{\rm (ml)}| \geq |\dot{\nu}^{\rm (obs)}|$ 
    \item RLO accretion with Kd structure: $|K^{\rm (rlo)}| \geq |K^{\rm (kd)}| \geq |K^{\rm (obs)}| \, \Rightarrow \,  |\dot{\nu}^{\rm (w)}| \geq |\dot{\nu}^{\rm (kd)}| \geq |\dot{\nu}^{\rm (obs)}|$ 
\end{itemize}

\noindent
Here, $K^{\rm (obs)}$ is an estimate of the torque obtained from Eq.~\ref{1.4.1.1}, where $\dot{\nu}_{\rm s} = \dot{\nu}^{\rm (obs)}$ is a value of spin frequency changes obtained from observations.

It is also important to note that regardless of the accretion regime and structure, an NS in a massive binary system will tend to an equilibrium period, but the value of this period depends on the accretion flow structure being realized. The spin rotation of an accreting NS near its equilibrium period is accompanied by alternating chaotic processes of acceleration and deceleration (local spin-up and spin-down) on short time scales caused by minor fluctuations in the value of the equilibrium period, which in turn are caused by inhomogeneities in the captured matter. The equilibrium period's average value should not change within this approach's framework. However, this value changes towards spin-up for most sources on large time scales. Consequently, there must be a mechanism and energy source to implement the observed manifestations of NS spin acceleration. In the next Sections, we consider tidal interactions between an NS and its massive companion as a possible mechanism influencing the NS long-term spin evolution and providing its super-global spin-up.

\section{Orbital periods and orbital decay in HMXB pulsar systems} \label{sec:orb}
\subsection{Observable characteristics}
The orbital periods of X-ray pulsars in the Galactic population of HMXB systems range widely from $\sim1.125$ to $\sim 380$ days \citep{Kim...2023ApJS..268...21K}. Persistent sources predominantly have small orbital eccentricities $\varepsilon < 0.2$ \citep{Tutukov...2020PhyU...63..209T, Fortin...2023AA...671A.149F} and relatively short orbital periods $P_{\rm orb}< 20$ days. Comparatively to persistent pulsars, transients, on average, have longer orbital periods and larger orbital eccentricities $\varepsilon > 0.2$, reaching $\varepsilon \simeq 0.88$ in the GS~1843$-$02 system \citep{Finger...1999ApJ...517..449F}. 

Orbital periods are also not constant and undergo changes due to orbital decay. These changes in the orbital period of HMXB X-ray pulsars was first recorded in the 1970s. Using the UHURU space telescope, it was established that during the observation period of 1971--1975, the orbital period of the pulsar Cen~X-3 decreased at a rate of $\dot{P}_{\rm orb}/P_{\rm orb} \approx -8\times 10^{-6}\,\text{yr}^{-1}$ \citep{Fabbiano...1977ApJ...214..235F}. For at least 8 sources from the Galactic population of X-ray pulsars in HMXB, a quasi-monotonic decrease in the orbital period has been detected (see Table~\ref{tab:pulsar_decay}): Cen~X-3, 4U~1538$-$522, SAX~J1802.7$-$2017, XTE~J1855$-$026, Vela~X-1, EXO~1722$-$363, OAO~1657$-$415, GX~301$-$2 \citep{Falanga...2015AA...577A.130F, Manikantan...2024MNRAS.527..640M}. Unlike measurements of the evolution of the spin period (spin-ups / spin-downs) of pulsars, obtaining estimates of changes in the orbital period is more difficult since these changes occur on large time scales.

\begin{deluxetable}{llcl}
\tabletypesize{\scriptsize}
\tablewidth{0pt} 
\tablecaption{Estimates of orbital period change of X-ray pulsars in HMXBs and LMXB pulsar Her~X-1 for corresponding timescales in Modified Julian Date. \label{tab:pulsar_decay}}
\tablehead{
    \colhead{Object} & \colhead{$\dot{P}_{\rm orb}/P_{\rm orb}$} & \colhead{MJD} & \colhead{References}\\
\colhead{}& \colhead{($\times10^{-6}\,\text{yr}^{-1}$)} & \colhead{(days)} & \colhead{}}
\startdata 
        Cen~X-3   & $-1.800\pm0.001$ &  $50087$--$55747$ & \cite{Falanga...2015AA...577A.130F}\\
        OAO~1657$-$415  & $-3.4\pm0.1$   & $50091$--$55747$ & \cite{Falanga...2015AA...577A.130F}\\
        Vela~X-1  & $-0.1\pm0.3$ &  $50087$--$55748$ & \cite{Falanga...2015AA...577A.130F}\\
        4U~1538$-$52   & $-0.95\pm0.37$ & $41446$--$57611$ & \cite{Hemphill...2019ApJ...873...62H}\\
        GX~301$-$2  & $-17.4\pm2.5$  & $43906$--$57489$ & \cite{Manikantan...2024MNRAS.527..640M}\\
        \hline
        Her~X-1  & $-0.01\pm0.0003$  & $41329$--$54345$ & \cite{Staubert...2009AA...500..883S}\\
      \enddata
\end{deluxetable}

One possible mechanism explaining the orbital decay of a high-mass X-ray binary system is the tidal interaction between the system's components \citep{Zahn...1977AA....57..383Z} caused by the gravitational interaction between an NS and its massive companion. In the next Subsection, we consider this phenomenon.

\subsection{Tidal interactions}
\label{sub:tides}
Tidal interactions in close binary systems are usually of two principal kinds: equilibrium tides (ET) and dynamical tides (DT). Equilibrium tides produce quasi‑static stellar deformations (``tidal bulges'') whose long axes align approximately with the line connecting the centers of two stars. In stars that possess radiative cores and convective envelopes, the tidal energy is dissipated mainly by turbulent viscosity in the envelope, making the equilibrium‑tide mechanism particularly efficient \citep{Zahn...1977AA....57..383Z}. By contrast, dynamical tides dominate in stars with convective cores and radiative envelopes, such as the early‑type massive companions in the high‑mass X‑ray binaries considered here. Because the companions of the neutron stars in our sample are early‑type massive stars with convective cores and radiative envelopes, the analysis below is framed entirely in terms of dynamical tide theory.

The framework of DT theory was introduced by \citet{Zahn...1970AA.....4..452Z, Zahn...1977AA....57..383Z}, who showed that the key ingredients in close binaries are the star’s normal-mode eigenfrequencies, which can resonate with the external tidal potential and thereby modify the binary’s orbital parameters.  For a non-degenerate star, the relevant eigenfrequencies fall in the range $0.1 \lesssim \sigma \lesssim 1$, where  $\sigma=\omega_{\rm f}/\sqrt{GM_{\rm opt}/R_{\rm opt}^{3}}$ is the dimensionless mode frequency and $\omega_{\rm f}$ is the physical frequency of the oscillation; this interval corresponds to high-order gravity (g-) modes \citep{Cowling...1941MNRAS.101..367C}. The spectrum of g-modes for non-degenerate stars is dense enough that the distance between adjacent frequencies can be significantly smaller than the frequency itself: $\omega_{\rm j} + \omega_{\rm j+1} << \omega_{\rm j}$. Here, the boundary between the convective core of the star and its radiative envelope plays an important role, since the main interaction of g-modes and tidal potential occurs in this boundary layer \citep{Zahn...1977AA....57..383Z, Chernov...2017AstL...43..429C}.

In the scenario of DT interaction, as the NS moves along its orbit, it can induce oscillations in the companion star, particularly near the gravitational (g-mode) resonant frequency. This phenomenon occurs if the orbital frequency of the HMXB system matches one of the oscillation frequencies (or its harmonics) of a massive star. Tidal interaction transfers orbital angular momentum to the spin angular momentum of the massive companion, ultimately leading to the orbital decay of the HMXB system and driving the system toward synchronous rotation, where the massive companion’s spin matches the orbital period \citep{Witte...1999AA...350..129W, Chernov...2020ARep...64..425C}. Some of the spin angular momentum of the massive star is, in turn, carried away by its ejected matter (stellar wind or flow through the interior Lagrange point) \citet{Chernov...2020ARep...64..425C}, which the NS partially captures. Thus, the complex gravitational interaction of a massive companion and an accreting NS in a binary system can provide a mechanism for transferring orbital angular momentum to the spin angular momentum of the NS. Next, we will consider the estimates obtained within the framework of the DT theory in more detail.

According to \citet{Witte...1999AA...350..129W, Witte...2001AA...366..840W}, a binary system has a reserve of orbital energy and angular momentum:
\begin{eqnarray}
    |E_{\rm orb}| &=&  \frac{G\, M_{\rm opt}\, M_{\rm ns}}{2a} \\ \nonumber
   |H_{\rm orb}| &=& \frac{M_{\rm opt}\, M_{\rm ns}}{M_{\rm opt} + M_{\rm ns}} a^{2}\, \Omega_{\rm orb} \sqrt{1 - e^2}
\end{eqnarray}
where $M_{\rm ns}$ is the mass of a NS, $M_{\rm opt}$ is the mass of an optical companion (massive component), $e$ is an orbital eccentricity, and $a$ is a semimajor axis of the orbit.
As a neutron star orbits, its perturbing tidal potential creates components of torque applied to its massive companion:
\begin{equation}
\label{eq:k_orb}
    K^{\rm \ell m}_{\rm n} = -\pi c^{\rm \ell m}_{\rm n} \int_{0}^{R}\int_{0}^{\pi} Im[\rho'(r, \upsilon)]\, P_{\rm \ell}^{\rm m}\, \cos{\upsilon} \, r^{\rm \ell+2}\, \sin{\upsilon}\, d\upsilon dr
\end{equation}
Here $\ell, \,m$ is tidal spherical harmonic components, $n$ is series of harmonics, $P_{\rm \ell}^{\rm m}$ and $c^{\rm \ell m}_{\rm n}$ are Hansen coefficients, $Im[\rho'(r, \upsilon)]$ is an imaginary part of the parameter of tidal perturbation of the stellar mass density. The torque leads to changes in the spin rotation of the massive companion, as well as to the evolution of the orbital energy and angular momentum of the binary system on the timescale:
\begin{eqnarray}
\label{eq:tidal_evol}
    \dot{E}_{\rm orb} &=&  - \sum_{\rm \ell , m} \sum_{\rm n} \dot{E}^{\rm \ell m}_{\rm n} \quad \text{where:} \quad \dot{E}^{\rm \ell m}_{\rm n} = \sigma_{\rm n}  K^{\rm \ell m}_{\rm n} \\ \nonumber
   \dot{H}_{\rm orb} &=& - \sum_{\rm \ell , m} \sum_{\rm n} \dot{H}^{\rm \ell m}_{\rm n} \quad \text{where:} \quad \dot{H}^{\rm \ell m}_{\rm n} = m  K^{\rm \ell m}_{\rm n}
\end{eqnarray}

According to \citet{Ivanov...2013MNRAS.432.2339I} and \citet{Chernov...2020ARep...64..425C}, the analytical solution of the equation \ref{eq:tidal_evol} for the energy loss in orbital motion due to the dynamical tidal interaction can be expressed by the following formula:

\begin{equation}
\dot{E}_{\rm orb} = \frac{\dot{P}_{\rm orb}}{P_{\rm orb}} \times \frac{G\, M_{\rm ns}\, M_{\rm opt}}{3a},
\label{eq:dot_e_orb}
\end{equation}
where $\dot{P}_{\rm orb}/P_{\rm orb}$ is a parameter of orbital period change.

In Table~\ref{tab:pulsar_energy_loss}, column $|\dot{E}_{\rm orb}|$, absolute values of the energy loss of the system due to orbital decay are given according to Eq.~\ref{eq:dot_e_orb}. The parameters of the pulsars used for the calculations in Table~\ref{tab:pulsar_energy_loss} are given in Table~\ref{tab:pulsar_opt} and Table~\ref{tab:pulsar_mass_comp}, and the cited literature (according to the references). Further discussion and interpretation of the results obtained from Eq.~\ref{eq:k_orb}--\ref{eq:dot_e_orb} is given in Section \ref{sec:disc}. In the next Section~\ref{sub_massive_comp}, we consider the influence of tidal interactions on the rotational evolution of massive components in HMXB systems.

\begin{deluxetable}{lrrrcrc}
\tabletypesize{\scriptsize}
\tablewidth{0pt} 
\tablecaption{Parameters of selected HMXB X-ray pulsars and LMXB pulsar Her~X-1. Spin periods $P_{\rm s}$ are rounded to the second decimal place. The ranges and average X-ray luminosity of sources $L_{\rm x}$ corresponds to the $18$--$50$~keV energy range obtained by INTEGRAL during 14 years of observations. The masses of NSs ($M_{\rm ns}$) are given in solar mass units ($M_\odot$), a semimajor axis of the orbit $a$ is given in astronomical units (AU), estimated using the third Kepler's law.}
\tablehead{
    \colhead{Object} & \colhead{$P_{\rm s}$} & \colhead{$P_{\rm orb}$} & \colhead{$L_{\rm x}$} & \colhead{$L_{\rm x}$ (aver)} & \colhead{$M_{\rm ns}$} & \colhead{$a$} \\
\colhead{} &\colhead{(s)} &\colhead{(days)} &\colhead{(erg s$^{-1}$)} &\colhead{(erg s$^{-1}$)} & \colhead{($M_\odot$)} &  \colhead{(AU)}}
\startdata 
        Cen~X-3 & 4.80 [1]& 2.03 [1] & (0.058--1.4)E37 [9] & 4.0E36 [9] & 1.34$\pm$0.16 [10] & 0.09 \\
        OAO~1657$-$415 & 37.02 [2]& 10.45 [6] & (0.1--2)E37 [9] & 5.8E36 [9]& 1.42$\pm$0.26 [11] & 0.23  \\
        Vela~X-1 & 283.43 [3]& 8.96 [7]& (0.0061--1)E37 [9] & 1.3E36 [9]& 2.12$\pm$0.16 [12] &  0.26  \\
        4U~1538$-$52 & 526.41 [4]& 3.73 [4] & (0.42--4.3)E36 [9] & 9.2E35 [9]& 1.18$\pm$0.29 [13] &  0.13 \\
        GX~301$-$2  & 672.51 [5]& 41.51 [8] & (0.023--3)E37 [9] &  2.8E36 [9]& 1.4? & 0.83   \\
        \hline
        Her~X-1  & 1.24 [14] & 1.70 [14] & (0.99--3.34)E37 [15] & 2.5E37 [16] & 1.5$\pm$0.3 [17] & 0.04   \\
\enddata
\tablecomments{[1] \cite{Shirke...2021JApA...42...58S}, [2] \cite{Sharma...2022MNRAS.509.5747S}, [3] \cite{Furst...2014ApJ...780..133F}, [4] \cite{Hemphill...2019ApJ...873...62H}, [5] \cite{Nabizadeh...2019AA...629A.101N}, [6] \cite{Jenke...2012ApJ...759..124J}, [7] \cite{Quaintrell...2003AA...401..313Q}, [8] \cite{Sato...1986ApJ...304..241S}, [9] \cite{Sidoli...2018MNRAS.481.2779S}, [10] \cite{Meer...2007AA...473..523V}, [11] \cite{Mason...2012MNRAS.422..199M}, [12] \cite{Falanga...2015AA...577A.130F}, [13] \cite{Abubekerov...2004ARep...48...89A}, [14] \cite{Staubert...2009AA...500..883S}, [15] \cite{Kosec...2020MNRAS.491.3730K}, [17] \cite{McCray...1982ApJ...262..301M}, [17] \cite{Reynolds...1997MNRAS.288...43R}.}
\label{tab:pulsar_opt}
\end{deluxetable}

\begin{deluxetable}{lrrrrrr}
\tabletypesize{\scriptsize}
\tablewidth{0pt} 
\tablecaption{Parameters of optical components of selected HMXB X-ray pulsars and LMXB pulsar Her~X-1. $M_{\rm opt}$ is the mass of optical component in solar masses ($M_\odot$), $R_{\rm opt}$ is the radius in solar radii ($R_\odot$), $\upsilon_{\rm rot} \sin{i}$ is the projected rotational velocity,  $P_{\rm rot}$ is the spin period in days, $P_{\rm ps}$ is the pseudosynchronisation period}
\tablehead{
    \colhead{Object} & \colhead{$M_{\rm opt}$} & \colhead{$R_{\rm opt}$} & \colhead{Opt. comp} &  \colhead{$\upsilon_{\rm rot} \sin{i}$} & \colhead{$P_{\rm rot}$} & \colhead{$P_{\rm ps}$}\\
\colhead{} & \colhead{($M_\odot$)} & \colhead{($R_{\odot}$)} & \colhead{Spec. class} &\colhead{(km s$^{-1}$)} & \colhead{(d)} & \colhead{(d)} }
\startdata 
        Cen~X-3 & 20.2$\pm$1.8 [1]& 12.1$\pm$0.5 [3] & O6.5 II--III [6] & 200$\pm$40 [3] & 2.54--3.39 [3] & $\sim$2.09 [3]\\
        OAO~1657$-$415 & 14.3$\pm$0.8 [1]& 25$\pm$2 [4] &  Ofpe/WNL [7]  & unknown  & unknown & $\sim$9.78 [*]\\
        Vela~X-1 & 26$\pm$1 [1]& 30.4$\pm$1.6 [3] &   B0.5 Ia [8] & 116$\pm$6 [3] & 11.59--14.29 [3] & 8.54--8.56 [3]\\
        4U~1538$-$52 &  $\sim$20 [1]& 13$\pm$1 [4]  &   B0.2 Ia [9] & 180$\pm$30 [4]  & 2.88--4.04 [*] & $\sim$3.12 [*]\\
        GX~301$-$2  & 43$\pm$10 [1]& $\sim$ 62 [3] &  B1.5 Ia [10] &  $\sim$50 [3]  & 48.08--58.98 [3] &  16.03--17.54 [3]\\
        \hline
        Her~X-1  & 2.3$\pm$0.3 [2] & 4.2$\pm$0.2 [5] & A7 V [11]  &  115$\pm$12 [5] & 1.67--2.06 [*] & $\sim$1.7 [*]\\
\enddata
\tablecomments{[1] - \cite{Fortin...2023AA...671A.149F}, [2] - \cite{Fortin...2024AA...684A.124F}, [3] - \cite{Stoyanov...2009AN....330..727S}, [4] - \cite{Falanga...2015AA...577A.130F}, [5] - \cite{Reynolds...1997MNRAS.288...43R},  [6] - \cite{Hutchings...1979ApJ...229..P1079}, [7] - \cite{Mason...2009AA...505..281M}, [8] -\cite{Nagase...1986PASJ...38..547N}, [9] - \cite{Parkes...1978MNRAS.184P..73P}, [10] - \cite{Kaper...1995AA...300..446K}, [11] - \cite{Leahy...2014ApJ...793...79L}, [*] - our calculations by using methods described in \cite{Stoyanov...2009AN....330..727S}}
\label{tab:pulsar_mass_comp}
\end{deluxetable}

\subsubsection{Spin period evolution of massive components}
\label{sub_massive_comp}
Tidal interactions shape the rotational (spin) evolution of the massive companion in an HMXB system: the orbital energy and angular momentum (see Eq.~\ref{eq:tidal_evol}) extracted by tides are transferred to the star’s spin, driving the system toward synchronisation of the stellar rotation and the orbital motion. In case of circular orbit ($ e\sim0$), the spin period of the massive component tends to reach the orbital period value $P_{\rm rot} \rightarrow P_{\rm orb}$ (equilibrium state) to minimise the influence of external tidal disturbances. If the orbit is eccentric, then equilibrium is achieved when $P_{\rm rot}$ reaches the so-called pseudosynchronisation period ($P_{\rm rot} \rightarrow P_{\rm ps}$), which is functionally related to the orbital period and eccentricity $e$ \citep{Hut...1981AA....99..126H, Stoyanov...2009AN....330..727S}:
\begin{equation}
\label{eq:ps_p}
    P_{\rm ps} = \frac{(1+3e^2 + \frac{3}{4}e^4)(1 - e^2)^{3/2}}{1 + \frac{15}{2}e^2 + \frac{45}{8}e^{4} + \frac{5}{16}e^6}P_{\rm orb}
\end{equation}

The article by \citet{Stoyanov...2009AN....330..727S} provides a comparative analysis of the estimates $P_{\rm ps}$ (according to Eq.~\ref{eq:ps_p}) and $P_{\rm rot}$ for sources: Cen~X-3, Vela~X-1, GX~301-2, where the spin period of massive component was obtained from expression: $P_{\rm rot} = 2\pi R_{\rm opt} \,\sin{i} / \upsilon_{\rm rot}\, \sin{i}$. Here: $R_{\rm opt}$ is the radius of a massive companion, $\upsilon_{\rm rot} \sin{i}$ is the projected rotational (spin) velocity, $i$ is the inclination of the orbit to the line of sight. In Table 4, in columns $P_{\rm rot}$ and $P_{\rm ps}$, we present the results of their calculations, as well as our results for objects OAO~1657-415 ($P_{\rm ps}$ only), 4U~1538-52 and Her~X-1. As can be seen from the Table, the massive components in the sources: Cen~X-3, Vela~X-1, GX~301-2 have not yet reached their synchronization (pseudosynchronization) periods and therefore must be in the phase of spin-up of axial rotation, 4U~1538-52, and Her~X-1 are close to these values. However, the observed negative dynamics of the orbital period change ($\dot{P}_{\rm orb} < 0$) also leads to a decrease in the synchronization period, which leads the normal components of the systems to forced spin-up. Thus, the systems we are considering have apparently not yet reached a state of tidal equilibrium.

In the following Sections, we consider possible relationships between the orbital decay phenomenon and super-global spin-up trends of X-ray pulsars in HMXB systems.

\section{Selected X-ray pulsars} \label{sec:select}

\textit{Cen~X-3} was discovered in 1967 using the Uhuru space observatory as a point X-ray source \citep{Chodil...1967PhRvL..19..681C}. Further observations revealed regular pulsations in its X-ray emission with a period of $P_{\rm s} \simeq 4.876 \, \text{s}$ \citep{Giacconi...1971ApJ...167L..67G}, which allowed the identification of this source as an X-ray pulsar. The detection of sinusoidal amplitude variations with a period of approximately $\sim 2.087 \,\text{days}$ led to the conclusion that Cen~X-3 is an eclipsing binary system with a massive companion \citep{Schreier...1972ApJ...172L..79S}. The eccentricity of the compact source's orbit is characterized by an extremely small value of $e \simeq 0.00021$ \citep{Fortin...2023AA...671A.149F}, close to a perfectly circular orbit. During observations from 2002 to 2016 using the INTEGRAL space observatory, the X-ray luminosity of the source in the range of $18$--$50$~keV smoothly varied within $L_{\rm x} \sim (0.058$--$1.4) \times 10^{37}\,\text{erg s}^{-1}$, with its average X-ray luminosity in a quiescent state being $L_{\rm x} \simeq 4 \times 10^{36}\,\text{erg s}^{-1}$. Optical observations in the vicinity of Cen~X-3 using various telescopes from the ESO and Las Campanas observatories (Chile) during 1973--1974 enabled the identification of the optical component of the system as an early spectral type star, 09--09.5~V or B0~I--III. Thus, Cen~X-3 became the first object identified as an X-ray pulsar in a massive binary system. Further studies of the source in the optical range refined the spectral class of the normal component to O6.5~II--III \citep{Hutchings...1979ApJ...229..P1079}, with a mass of approximately $\sim 20.2\, M_{\odot}$ \citep{Ash...1999MNRAS.307..357A}, located at a distance of $d \simeq 4.78 \,\text{kpc}$ \citep{Gaia...2023AA...674A...1G}. The optical component  (massive companion) in the Cen X-3 system is characterized by filling its Roche lobe with outflow of matter through the inner Lagrange point (L1) into the Roche lobe of the NS. The pulsar Centaurus X-3 (also SMC X-1, LMC X-4 and possibly OAO~1657$-$415) is one of the rare cases where the RLO accretion regime is realized in HMXB systems \citep{Fornasini...2023hxga.book..143F}. As can be seen from Fig.~\ref{fig:image1}, this pulsar has been showing a quasi-stable trend of spin-up since its discovery.
\\

\textit{OAO~1657$-$415} was initially discovered in 1978 as an eclipsing X-ray source \citep{Polidan...1978IAUC.3234....1P}, was identified a year later as an accreting X-ray pulsar, with a period of $P_{\rm s} \simeq 38.22$ seconds at that time \citep{White...1979ApJ...233L.121W}. The NS currently rotates with a period of $P_{\rm s} \simeq 37.02\, \text{s}$ and has a magnetic field strength on its surface, measured from the X-ray spectrum cyclotron line analysis, $B_{\rm ns} = (3.29 \pm 0.23) \times 10^{12}\,\text{G}$ \citep{Sharma...2022MNRAS.509.5747S}. The X-ray luminosity of the source, observed from 2002 to 2016 in the $18$--$50$~keV range, varied smoothly in the range of $L_{\rm x}\simeq(0.1$--$2) \times 10^{37} \,\text{erg s}^{-1}$ \citep{Sidoli...2018MNRAS.481.2779S}. OAO~1657$-$415 is located at a distance of $d \simeq 1.59 \,\text{kpc}$ \citep{Gaia...2023AA...674A...1G} and is identified as a massive X-ray binary system with an orbital period of $P_{\rm orb} \simeq 10.45 \,\text{d}$ and an eccentricity of $e \simeq 0.104$ \citep{Chakrabarty...2002ApJ...573..P789}. The components of the system are an NS and a massive supergiant of spectral class Ofpe. \citet{Baykal...1997A&A...319..515B} analysed the spin evolution of this pulsar and showed that OAO~1657$-$415 may be a wind accretor with the formation of a Keplerian accretion disk. In a further work \citet{Baykal...2011A&A...529A...7I} used the analysis of the power density spectra of X-ray sources. The authors showed that OAO~1657$-$415 may be in the RLO-accretion regime. This source has also been considered as a wind accretor with MAD-disk formation \citep{Kim...2017JPhCS.929a2005K}. The difficulty in interpreting this source is due to the poor study of its massive companion. As can be seen from Fig.~\ref{fig:image1}, this pulsar demonstrated a quasi-stable global trend of spin-up from the time of its discovery (1978) until approximately the 2010s with an average rate of $\dot{\nu}_{\rm su} \,
\simeq \, (8.3-8.9) \times 10^{-13} \text{Hz s}^{-1}$ \citep{Barnstedt...2008A&A...486..293B, Jenke...2012ApJ...759..124J}, after which the pulsar entered a stage of global spin-down.\\

\textit{Vela~X-1} was initially discovered by the Uhuru Space Observatory at the dawn of X-ray astronomy in 1967 \citep{Chodil...1967PhRvL..19..681C}. It was subsequently identified as an X-ray pulsar in a massive binary system. The spin period of the compact source is $P_{\rm s} \simeq 283.43 \,\text{s}$. Between 2002 and 2016, observations with the Integral space observatory revealed that the X-ray luminosity of the source in the $18$--$50$~keV range varied within $10^{34}$--$10^{37}\,\text{erg s}^{-1}$, while its average X-ray luminosity in a quiescent state was $L_{\rm x} \simeq 1.3 \times 10^{36}\,\text{erg s}^{-1}$ \citep{Sidoli...2018MNRAS.481.2779S}. The orbit has a low eccentricity, close to circular, with $ e\sim0.088$. The orbital period is $P_{\rm orb} \simeq8.96$ days. The companion of the NS is the supergiant star HD~77581, located at a distance of $d \simeq 1.68 \,\text{kpc}$ \citep{Gaia...2023AA...674A...1G}, belonging to spectral class B0.5~Ia, with an estimated mass of $M_{\rm opt} \sim 26\, M_{\odot}$. Based on spectral studies in the near-infrared, visible, ultraviolet ranges and also mathematical modeling,  the terminal wind velocity is estimated at $\upsilon_{\infty} \sim 600-700\,\text{km s}^{-1}$ and the approximate mass-loss rate is $\dot{M}_{\rm out} \sim 10^{-6}\, M_{\odot}\,\text{yr}^{-1}$ \citep{Gimenez-Garcia...2016AA...591A..26G, Sander...2018AA...610A..60S}. The Vela~X-1 pulsar is a wind accretor, the structure of which is considered in the framework of quasi-spherical subsonic accretion \citep{Shakura...2013PhyU...56..321S}, as well as in the framework of accretion from prograde \citep{Mellah...2019AA...622A.189E} and retrograde Keplerian disk \citep{Malacaria...2020ApJ...896...90M} and also MAD-disk \citep{Ikhsanov...2014ARep...58..376I}. As can be seen in Fig.~\ref{fig:image1}, the spin evolution of the pulsar is characterised by intense local alternating variations of spin-up and spin-down; according to \citet{Malacaria...2020ApJ...896...90M} these episodes have no clear correlation with observed X-ray flux. Unlike Cen~X-3, 4U~1538$-$52 and OAO~1657$-$415, this pulsar does not have quasi-stable global trends. However, on the full-time scale of its observations, it is clear that the pulsar has a general tendency to spin down. It is worth noting that the estimate of the orbital period change, $\dot{P}_{\rm orb}/P_{\rm orb} = 0.1 \pm 0.3$ (see Table~\ref{tab:pulsar_decay} in \citet{Falanga...2015AA...577A.130F}), for Vela~X-1 is smaller than the associated measurement uncertainty. This suggests that the obtained value is a qualitative rather than a quantitative indication of orbital decay.\\

\textit{4U~1538$-$52} was discovered in 1976 using the OSO-8 space observatory as an eclipsing X-ray source in a binary system. Further studies during the Ariel-5 space mission revealed pulsations in the X-ray flux with a period of approximately $\sim 528.93\, \text{s}$ associated with the spin rotation of the NS \citep{Davison...1977MNRAS.181P..73D}. As of the current epoch, the spin period of 4U~1538$-$52 corresponds to $P_{\rm s} \simeq 526.23 \,\text{s}$ \citep{Tamang...2024MNRAS.527.3164T}, and the orbital period is $P_{\rm orb} \simeq 3.728 \,\text{d}$. During observations from 2002 to 2016 using the Integral space observatory, the X-ray luminosity of the source in the range of $18$--$50$~keV smoothly varied within $L_{\rm x} \sim (0.42$--$4.3) \times 10^{36}\,\text{erg s}^{-1}$ \citep{Sidoli...2018MNRAS.481.2779S}. Spectral observations in the optical range using the 3.9-meter telescope (Siding Spring Observatory, Australia) identified the normal companion of the system (QV~Nor) as a supergiant star of early spectral type B0.2~Ia \citep{Parkes...1978MNRAS.184P..73P} with a mass-loss rate of approximately $\sim 10^{-6}\, M_{\odot}\,\text{yr}^{-1}$. The distance to the source is estimated to be $d \simeq 6.81 \,\text{kpc}$ \citep{Gaia...2023AA...674A...1G}. The pulsar 4U~1538$-$52 is a wind accretor that exhibits alternating global trends of spin-up and spin-down, the duration of which is about $1$--$2$~decades with rates $\sim 10^{-14}\,\text{Hz s}^{-1}$ \citep{Hu...2024ApJ...971..120H, Sharma...2023MNRAS.522.5608S}. The estimate of the orbital period change $\dot{P}_{\rm orb}/P_{\rm orb} = 0.95 \pm 0.37$  (see Table~\ref{tab:pulsar_decay}, and \citet{Hemphill...2019ApJ...873...62H}) has a rather large uncertainty (the statistical significance of its deviation from zero is about \(2.6\sigma\)). This implies that with about $95\%$ confidence, orbital decay in the system 4U~1538$-$52 is indeed present, but more detailed studies are needed to obtain more reliable estimates. \\

\textit{GX~301$-$2} is a unique object among the population of persistent X-ray pulsars in HMXB systems, possessing an orbit with a significant eccentricity of $e \simeq 0.46$ \citep{Fortin...2023AA...671A.149F}, which is more typical for transient X-ray pulsars. In most cases, persistent pulsars have an eccentricity not exceeding $0.2$. Between 2002 and 2016, observations with the Integral space observatory revealed that the X-ray luminosity of the source in the $18$--$50$~keV range varied within $(0.023$--$3)\times10^{37}\,\text{erg s}^{-1}$, while its average X-ray luminosity in a quiescent state was $L_{\rm x} \simeq 2.8 \times 10^{36}\,\text{erg s}^{-1}$ \citep{Sidoli...2018MNRAS.481.2779S}. The optical component of the system is the supergiant star BP~Cru (Wray~977) of early spectral type B1.5~Ia, with a mass of $M_{\rm opt} \sim 48 \, M_{\odot}$, the largest among the entire galactic population of X-ray pulsars in massive binary systems. The estimated distance to the source is $3.5 \pm 0.5$~kpc \citep{Kaper...1995AA...300..446K}. Spectral studies of the optical component using the 3.6-meter ESO telescope (La Silla Observatory, Chile) revealed the presence of a P~Cygni profile, which allowed estimates of the mass-loss rate to be $\dot{M}_{\rm out} \simeq (0.5$--$1)\times 10^{-5}\, M_{\odot}\,\text{yr}^{-1}$ with a terminal wind velocity of $\upsilon_{\infty} \sim 400\,\text{km s}^{-1}$ \citep{Kaper...1995AA...300..446K}. The spin evolution of GX~301$-$2, likewise of Vela X-1, is chaotic, characterized by intense global and local trends of varying durations, reaching an absolute value of  $\sim 6\times10^{-12}\,\text{Hz s}^{-1}$ \citep{Malacaria...2020ApJ...896...90M}. Since GX~301$-$2 is a wind accretor, its peculiar spin evolution can be explained by the formation of hybrid (variable) accretion structures surrounding the NS, driven by changes in the accretion flow parameters. In such scenarios, the quasi-spherical flow can transform into an accretion disk and vice versa \citep{Nabizadeh...2019AA...629A.101N}. These transitions predominantly occur in HMXB systems with a Be-type companion star (see Subsection~\ref{sub:regimes}).

\section{Discussion} \label{sec:disc}

\subsection{Linking Observations to Theory}
In Section~\ref{sec:spin} we examined the spin evolution of NSs, specifically those that manifest as X-ray pulsars in massive binary systems with the longest history of observations. As shown in Fig.~\ref{fig:image1}, the spin evolution of these objects displays a complex structure consisting of multiple components, including local and global trends of spin-up and spin-down. Notably, all objects in the sample (except Vela~X-1) exhibit a general trend of accelerating spin rotation over several decades, from their discovery to the present day. Estimates of these super-global trends for X-ray pulsars, derived from the least squares method using a linear approximation over the specified time intervals, are presented in Table~\ref{tab:pulsar_nu_dot_super}. It is known that the energy per unit time needed for changes in the spin rotation of NS is expressed as:

\begin{equation}
    \dot{E}_{\rm s} = I\,\omega_{\rm s}\, \dot{\omega}_{\rm s} =  -I\,\frac{4\pi^2\, \dot{\nu}_{\rm s}}{P_{\rm s}}
    \label{eq:es}
\end{equation}

\noindent
where $I$ is a moment of inertia of a NS. For our calculations we used the canonical value $I_{\rm ns} = 10^{45}\,\text{g cm}^2$, $\omega_{\rm s} = 2\pi/P_{\rm s}$ is the spin angular velocity of NS's, $\dot{\omega}_{\rm s}$ is its derivative.

The Eq.~\ref{eq:es} allows us to estimate the energy loss of the system spent on spinning up the axial rotation of the NS. Table~\ref{tab:pulsar_energy_loss}, column $|\dot{E}_{\rm s}|$, provides estimates of the energy losses under consideration, obtained using Eq.~\ref{eq:es}. Estimates of the $|\dot{E}_{\rm s}/\dot{E}_{\rm orb}|$ ratios for the X-ray pulsars in the sample are also provided. As can be seen from the table (in column $|\dot{E}_{\rm s}/\dot{E}_{\rm orb}|$), the energy losses due to orbital dissipation are sufficient ($|\dot{E}_{\rm s}/\dot{E}_{\rm orb}| \ll 1$) to ensure long-term spin-up of the pulsar at the observed rates. In Table~\ref{tab:pulsar_energy_loss} in the column $\dot{\nu}_{\rm s}^{\rm max}$ we showed maximal possible values of NS spin-up trends in the frame of DT theory, estimated using Eq.~\ref{eq:dot_e_orb} and Eq.~\ref{eq:es}: 

\begin{equation}
\label{eq:dnu_max_td}
\dot{\nu}_{\rm s}^{\rm max} \simeq 5.2\times 10^{-11}\,\text{Hz s}^{-1} \, \times \, I_{45}^{-1}\, m\, \left(\frac{P_{\rm s}}{37} \right)  \left( \frac{M_{\rm opt}}{14.3\, M_{\odot}}\right)  \left( \frac{\dot{P}_{\rm orb}/P_{\rm orb}}{-3.4 \times 10^{-6}\, \text{yr}^{-1}} \right)\left( \frac{a}{0.23\, \text{AU}} \right)^{-1}
\end{equation}

As seen from Table~\ref{tab:pulsar_nu_dot_super} and Table~\ref{tab:pulsar_energy_loss}, observable values of super-global spin-ups do not exceed their theoretical upper limits ($\dot{\nu}_{s}^{\rm max}\gg\dot{\nu}_{s}$). Moreover, in \citet{Witte...1999AA...341..842W} numerical estimates of the solution of integral Eq.~\ref{eq:k_orb} for a massive companion ($10 M_{\odot}$) are given; in this case, for possible $g_k$ modes of the components, $ K^{\rm \ell m}_{\rm n}$ lies in the range $10^{38}$--$10^{47}$ erg. Even at the minimum value $K^{\rm \ell m}_{\rm n}$, according to the Eq.~\ref{1.4.1.1}, it is sufficient to ensure NS spin period change rate $|\dot{\nu}_{\rm s}| \leq 10^{-8}\, \text{Hz s}^{-1}$, provided that the transfer of angular momentum occurs without losses.

In Section~\ref{sub_massive_comp} the rotational evolution of massive components in HMXB systems is considered, where it is shown that the sources (massive components) must be in the state of spin-up, since according to the estimates of $\dot{P}_{\rm orb}$, $P_{\rm ps}$ and $P_{\rm rot}$  they have not reached the state of tidal equilibrium. Thus, massive companions in HMXB systems can be transfer links, accepting energy from orbital dissipation for their own axial spin-up and then partially losing it through stellar wind (or matter flowing through the inner Lagrange point in the case of RLO-accretion). The subsequent interaction of the captured accretion flow with the neutron star at the boundary of its magnetosphere leads to the transfer of angular momentum, which in turn leads to the observed changes in the NS spin period. To sum up, we can make the following important conclusion: the regular super-global spin-up trends of X-ray pulsars in HMXB can be explained by the transfer of orbital angular momentum to the spin angular momentum through tidal interactions and subsequent orbital decay between the NS and its massive companion.

\begin{deluxetable}{lcccl}
\tabletypesize{\scriptsize}
\tablewidth{0pt} 
\tablecaption{Estimates of absolute values of energy loss $|\dot{E}_{\rm orb}|$ due to orbital decay, energy per unit of time needed for changes in the spin rotation of NS $|\dot{E}_{\rm s}|$ and their ratio  $|\dot{E}_{\rm s}/\dot{E}_{\rm orb}|$. Maximal possible values of NS spin-up trends $\dot{\nu}_{s}^{\rm max}$ in the model of angular momentum exchange due to the tidal interaction. \label{tab:pulsar_energy_loss}}
\tablehead{
\colhead{Object} & \colhead{$|\dot{E}_{\rm s}|$} & \colhead{$|\dot{E}_{\rm orb}|$} & \colhead{$|\dot{E}_{\rm s}/\dot{E}_{\rm orb}|$} & \colhead{$\dot{\nu}_{\rm s}^{\rm max}$} \\
\colhead{}& \colhead{(erg s$^{-1}$)} & \colhead{(erg s$^{-1}$)} & \colhead{} & \colhead{(Hz s$^{-1}$)} }
\startdata 
        Cen~X-3        & 9.51E33 & 1.05E35 & 0.09 & 1.27E-11 \\
        OAO~1657$-$415 & 5.43E32 & 5.54E34 & 0.01 & 5.20E-11 \\
        Vela~X-1       & 7.80E29 & 4.03E33 & 2E-4 & 2.90E-11 \\
        4U~1538$-$52   & 8.84E29 & 3.24E34 & 3E-5 & 4.32E-10 \\
        GX~301$-$2     & 1.76E30 & 2.45E35 & 7E-6 & 4.18E-9  \\
        \hline
        Her~X-1     & 1.40E33 & 1.49E32 & 9.35 & 4.69E-15  \\
      \enddata
\end{deluxetable}

In the special case of the spin evolution of Vela~X-1 (see Fig.~\ref{fig:image1}), a negative trend in the spin frequency is observed, unlike the other objects in the sample. Over the entire observation period, this pulsar demonstrates the spin-down process with a rate of $\dot{\nu}_{\rm s} \approx -5.6 \times 10^{-15}\, \text{Hz s}^{-1}$ (see Table~\ref{tab:pulsar_nu_dot_super}). This phenomenon can be explained by the fact that, probably, a larger time scale is required to observe the super-global trend of spin-up for Vela~X-1. As can be seen from Eq.~\ref{sa_qsp} and Eq.~\ref{nu_qsp_up}, the rate of spin frequency changes depends significantly on the relative velocity (which in turn depends on the wind velocity) $\dot{\nu}_{\rm s} \sim [\upsilon_{\rm rel}(\upsilon_{\rm w})]^{-4}$. Even a small fluctuation in $\upsilon_{\ rmw}$, therefore, produces a large change in the sign and magnitude of the torque, potentially dominating over the slower tidal contribution. We assume that on large time scales, this source, like the other pulsars in the sample, demonstrates a super-global trend of spin-up associated with the transfer of orbital angular momentum. Perhaps long-period variations in the wind of the massive companion in the Vela~X-1 system have periods greater than 50 years. In this case, alternating global trends of acceleration and deceleration of the Vela~X-1 pulsar form the structure of spin evolution with the corresponding long ($>50$~years) periodicity against the background of the super-global trend of spin-up. Thus, one of the global trends of spin-down falls on the current epoch.

\subsection{Comparison with X-ray pulsars in LMXB systems}
\label{sub_LMXB}
The Galactic population of known X-ray pulsars in low-mass X-ray binary (LMXB) systems includes only $33$~sources, of which $23$~objects belong to the subclass of millisecond pulsars \citep[MSPs,][]{Avakyan...2023A&A...675A.199A}. The vast majority of MSPs have weak magnetic fields ($\sim 10^{8}$--$10^{9}\, \text{G}$) and a non-accretion nature of their X-ray emission. The LMXB X-ray pulsars are mainly old systems \citep[$\geq 10\, \text{GYr}$,][]{Kar...2024MNRAS.535..344K}. On such time scales, the strong magnetic field of an NS has time to weaken significantly \citep{Igoshev...2021Univ....7..351I}. Among the population of accreting X-ray pulsars in LMXB systems, only two sources have been well-studied for a long time: Her~X-1 and GX~1+4. Of these two sources, a change in the orbital period was detected only in Her~X-1. Let us consider it in more detail.

\textit{Her~X-1} was originally discovered in 1972 by the UHURU space observatory as a compact object with a pulsation period of $1.24$~s in the Hercules constellation \citep{Tananbaum...1972ApJ...174L.143T}. Further observations established the source's binarity by analysing the Doppler shifts in the pulsation profiles with an orbital period of $P_{\rm orb} \simeq 1.7\,\text{d}$. The fast pulsations and the similarity with the observed characteristics of Cen~X-3 left no doubt that it was an accreting NS \citep{Tananbaum...1972ApJ...174L.143T}. Optical studies of Her~X-1 have indicated that the normal component of the system is a star HZ~Her of spectral type A7, it fills its Roche lobe \citep{Leahy...2014ApJ...793...79L} and has a low orbital eccentricity close to circular $e\simeq 0.0003$ \citep{Deeter...1981ApJ...247.1003D}. The distance to the source is approximately $d \sim 6 \,\text{kpc}$ \citep{Bahcall...1972ApJ...178L...1B}. The estimated mean X-ray luminosity at this distance is $L_{\rm x} \simeq 2.5 \times 10^{37}\, \text{erg s}^{-1}$ at range $2$--$60 \, \text{keV}$ \citep{McCray...1982ApJ...262..301M}. An analysis of the X-ray spectrum of Her~X-1 revealed the presence of a cyclotron line, from which an estimate of the magnetic field on the surface of the NS was obtained, $B_{\rm ns} \simeq 5.3\times10^{12}\, \text{G}$ \citep{Truemper...1978ApJ...219L.105T}. Since the first observations, it has been noted that this X-ray source has an additional long-period variation in brightness with a period of $\sim 35$ days, which also occurs in the optical range. One of the main hypotheses in interpreting this phenomenon is the assumption of a precessing deformable accretion disk around the NS. Quasi-periodic deformations in the disk are caused by the tidal effects of a massive companion \citep{Leahy...2011ApJ...736...74L}. Her~X-1 is a system where the flow of matter from the normal component to the NS occurs in the RLO-accretion regime with the formation of a Keplerian disk. As shown in Fig.~\ref{fig:image1}, the large-scale spin evolution of Her~X-1 is similar to that of the HMXB pulsar Cen~X-3. 

Since its discovery, Her~X-1 has also shown a quasi-stable superglobal spin-up trend estimated as $\sim 4.39\times 10^{-14}\,\text{Hz s}^{-1}$ (see Table~\ref{tab:pulsar_nu_dot_super}). However, as can be seen from Table~\ref{tab:pulsar_energy_loss}, the estimates of the energy loss due to the observed orbital decay of Her~X-1 are significantly less than the energy per unit time providing the observed super-global spin-up of the NS, $|\dot{E}_{\rm s}| >> |\dot{E}_{\rm orb}|$. 

The key difference between the RLO-accreting systems Cen X-3 and Her X-1 is that the optical component of the Cen X-3 system is a hot O6 II-III giant star with a convective core and a radiative envelope where dynamical tides are effective (see Section \ref{sub:tides}); while the optical component in the Her X-1 system is a later spectral type (A7 V) main sequence star with a radiative core and a convective envelope where dynamical tides are not effective \citep{Zahn...1977AA....57..383Z}. Thus, the mechanism of angular momentum exchange due to the tidal interaction is not the main one for ensuring the observed super-global spin-up of Her X-1.

\section{Conclusions} \label{sec:conc}
The spin evolution of NSs, manifesting as X-ray pulsars in HMXB systems, is a complex, multicomponent process driven by the interaction between the NS and its massive companion via captured matter. The structure of this evolution can be divided into several key components:

Local trends of spin-up and spin-down: These are episodic, chaotic variations in the NS's spin period, occurring at a high rate but lasting for relatively short durations ranging from several days to months. These fluctuations arise because the accreting NS rotates near its equilibrium period, and inhomogeneities in the captured matter cause small period deviations around this value. Local trends occur against the backdrop of broader global trends of spin-up and spin-down.

Global trends of spin-up and spin-down: These trends are approximately an order of magnitude smaller in absolute value compared to local trends but span longer time scales, from several months to years. Several hypotheses have been proposed to explain these phenomena, including asymmetry in the scalar potential from the accretion flow and long-period variations in the stellar wind of the massive companion. The maximum possible local and global rates of NS spin frequency changes correspond to the theoretical limits specified in Eq.~\ref{nu_qsp_up} and Eq.~\ref{nu_RLO}, depending on the realized accretion regime.

\begin{figure}
    \centering
    \includegraphics[width=0.3\columnwidth]{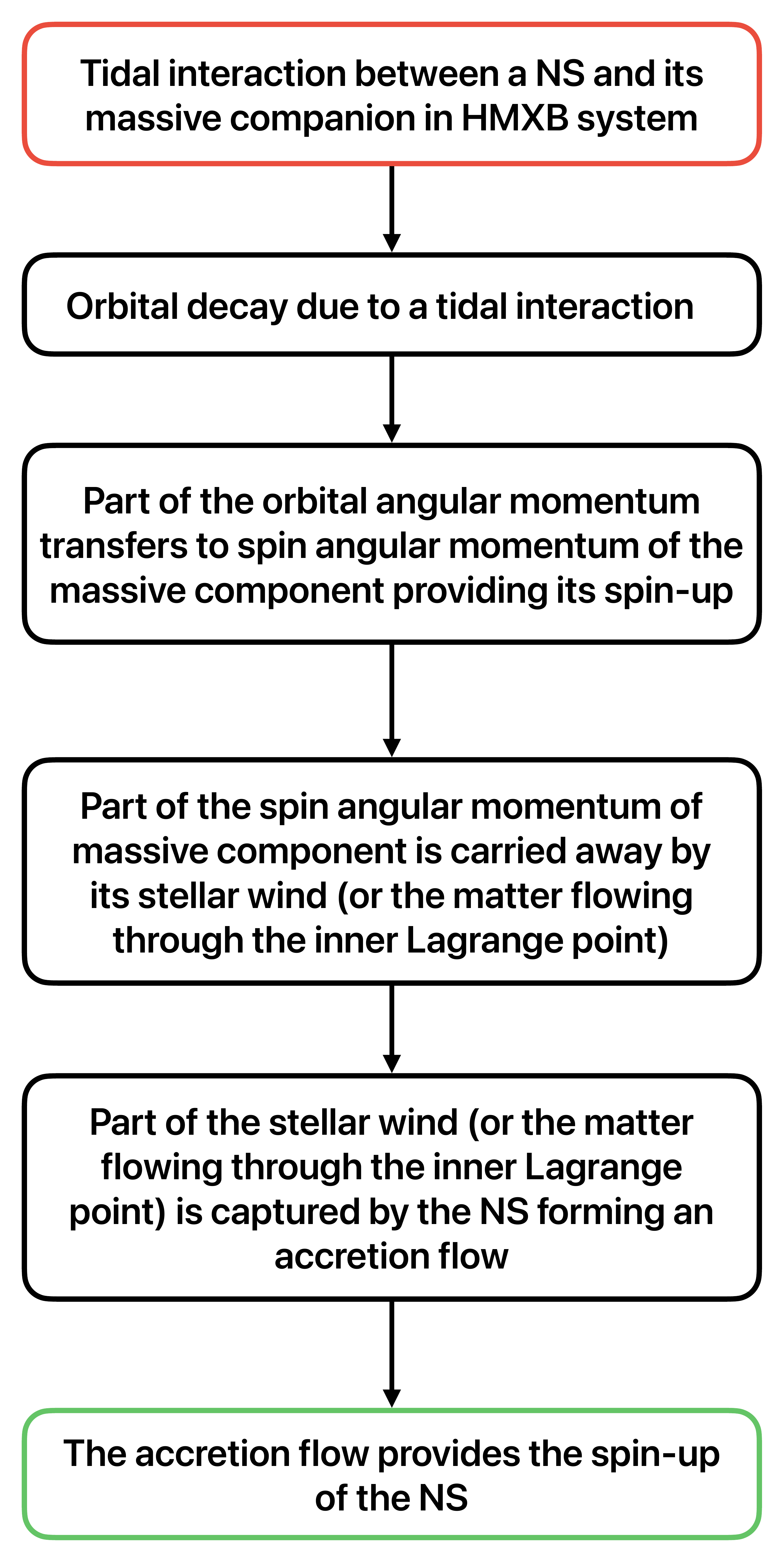}
    \caption{A simplified scheme showing the mechanism by which a high-mass X-ray binary system's orbital angular momentum is transferred to NS's spin angular momentum.}
    \label{fig:scheme}
\end{figure}

Super-global trends of spin-up: These trends represent the largest-scale changes in the spin period over the spin evolution of persistent X-ray pulsars in HMXB (RLO and sgXRB) systems. In this paper, we show that for the majority of sources with long-term observational records, a slow but steady spin-up of axial rotation is observed, with rates in range $\dot{\nu} \sim 10^{-12}$--$10^{-14}$~Hz~s$^{-1}$. We suggest that the primary mechanism driving these NS spin-up processes is the transfer of orbital to spin angular momentum. In this scenario, tidal torques drive orbital decay, transferring orbital energy and angular momentum to the massive companion and spinning it up. Part of that newly acquired spin angular momentum is removed by the stellar wind (or the matter flowing through the inner Lagrange point in case of RLO-accretion); a fraction of that outflow is then gravitationally captured by the neutron star and forms an accretion flow. The accretion flow’s own angular momentum is subsequently transferred to the neutron star, producing the observed long-term spin changes of the NS (see Fig.~\ref{fig:scheme}). The tidal-decay energy loss and angular momentum in HMXB (RLO and sgXRB) systems are sufficient to account for the observed long-term spin-up of X-ray pulsars.

The RLO-accretion regime is more efficient for transferring angular momentum from a massive companion to an NS compared to the wind accretion regime since, in this case, the capture radius of matter exceeds the Bondi radius, which creates a larger torque applied to the NS. However, even in the wind accretion regime, the torque is sufficient to provide the observed super-global spin-up trends.

The similarity of the super-global spin-up trends in the monotony of the HMXB pulsar Cen~X-3 and the LMXB pulsar Her~X-1, as well as the relative chaos of the observed trends of wind-fed HMXB pulsars, indicate that the spin evolution over long time intervals is determined primarily by the accretion regime.

It is shown that in the case of Her X-1, the energy losses from tidal interactions are not sufficient to provide the observed NS spin-up, see Table\ref{tab:pulsar_energy_loss}. Therefore, the main mechanism of its pulsar spin-up is not associated with the orbital decay of the system. The question still remains: is this typical for most of the accreting X-ray pulsars in LMXB systems or not? Due to the relatively small population of such sources in LMXB systems and the insufficient study of their spin and orbital evolution on long time scales (except for Her X-1), the question of the similarities and differences in spin evolution between HMXB and LMXB pulsars remains open.

\begin{acknowledgments}
This research is funded by the Science Committee of the Ministry of Science and Higher Education of the Republic of Kazakhstan under Grant No.~AP26103097, Grant No.~BR24992759,
and Grant No.~BR24992807.
\end{acknowledgments}

\vspace{5mm}
\facilities{Fermi \citep[GBM;][]{Fermi_pulsars}, CGRO \citep[BATSE;][]{Batse}, INTEGRAL \citep[][]{Integral_pulsars}.}

\software{\texttt{NumPy} \citep{Harris...Nature2020...585...7825H}, \texttt{pandas} \citep{Mckinney...PPSC2010...56M}, \texttt{Astropy} \citep{astropy:2013, astropy:2018, astropy:2022},  \texttt{Matplotlib} \citep{Hunter...CSE2007...9...90}, \texttt{Scikit-learn} \citep{scikit-learn2011}.}

\bibliography{sample631}{}
\bibliographystyle{aasjournal}

\end{document}